\documentclass[structabstract]{aa}  

\usepackage[pdftex]{graphicx}
\usepackage{natbib}
\usepackage[colorlinks=true,linkcolor=blue,anchorcolor=blue,citecolor=blue,urlcolor=blue]{hyperref}  

\begin{document}
\title{\object{PKS 1510-089} a rare example of a flat spectrum radio
  quasar with a very high-energy emission}

\author{Anna Barnacka\inst{1,2}\fnmsep\thanks{\email{abarnacka@cfa.harvard.edu}}
  \and Rafal Moderski \inst{1}  \and Bagmeet Behera
  \inst{4} \and Pierre Brun \inst{3} \and Stefan Wagner \inst{5} }

\institute{Nicolaus Copernicus Astronomical Center, Warsaw, Poland
  \and Harvard-Smithsonian Center for Astrophysics, Cambridge, MA 02138, USA
  \and CEA Irfu, Centre de Saclay, F-91191 Gif-sur-Yvette France
  \and Deutsches Elektronen-Synchrotron (DESY),Platanenallee 6, D-15738 Zeuthen, Germany
  \and Landessternwarte, Universitat Heidelberg, Konigstuhl, D-69117
  Heidelberg, Germany }


\abstract
    {The blazar PKS 1510-089 is an example of flat spectrum
      radio quasars. 
      High-energy emissions from this class of objects  are believed 
      to have been produced by inverse Compton radiation with 
      seed photons originating from the broad line region.     
    In such a paradigm, a lack of very high-energy emissions is expected 
    because of the Klein-Nishina effect and strong absorption in the
    broad line region.  Recent detection of at least three such
    blazars by Cherenkov telescopes has forced a revision of our
    understanding of these objects.
      }
    {We have aimed to model the observed spectral energy distribution
      of PKS 1510-089 from the high-energy flares in March 2009, 
      during  which very high-energy emission were also detected
      by H.E.S.S.}
   {We have applied the single-zone internal shock scenario to
     reproduce the multiwavelength spectrum of PKS~1510-089.  We have
     followed the evolution of the electrons as they propagate along
     the jet and emit synchrotron and inverse Compton radiation. We
     have considered two sources of external photons: the dusty
     torus and the broad line region.  We have also examined the
     effects of the gamma-gamma absorption of the high-energy photons both in 
     the AGN environment (the broad line region and the dusty torus),
     as well as while traveling over cosmological distances: the extragalactic background light.}
   {We have successfully modeled the observed spectrum of PKS
     1510-089. In our model, the highest energy emission is the result
     of the Comptonization of the infrared photons from the dusty
     torus, thus avoiding Klein-Nishina regime, while the bulk of the
     emissions in the GeV range may still  be dominated by the
     Comptonization of radiation coming from the broad line region. }
   {}

\authorrunning{Barnacka et al. 2013}
\titlerunning{PKS 1510-089 - FSRQ with VHE emission}
\maketitle
%

\section{Introduction \label{sec:intro}}

The observations by the {\it Fermi}/LAT instrument in the high-energy (HE: 100~MeV~$< E <$~100~GeV) 
range resulted in an identification of 1873 sources \citep{2012ApJS..199...31N}.
Most of these sources are blazars, which are very luminous,
active galactic nuclei (AGNs) with relativistic jets pointing toward the observer.

The broadband spectrum of blazars is dominated by non-thermal emissions
produced in a relativistic jet \citep{1978bllo.conf..328B}.  The
spectral energy distribution (SED) of blazars is characterized by two,
broad spectral components.  
A low-energy component extends from the
radio to optical/UV/X-rays, and  is produced by the synchrotron radiation of relativistic electrons.  
The high-energy component may extend from
X-rays to $\gamma$ rays and, according to recent
interpretations, is produced by inverse Compton (IC) radiation with a
possible source in seed photons, being either the synchrotron radiation,
the broad line region (BLR), or the dusty torus (DT).

Blazars can roughly be divided into two classes: flat spectrum radio
quasars (FSRQs) and BL Lac objects;  FSRQs are distinguished by the presence of  broad emission lines, 
which are absent or very weak in BL Lac objects.  

The high-energy component of FSRQs is usually
much more luminous than the low-energy one.
The seed photons for inverse Compton radiation most probably come
 from regions external to the jet \citep{1994ApJ...421..153S,2000ApJ...545..107B}.  
 The high-energy component of BL Lac objects results from the Comptonization of
synchrotron photons. 
The luminosity at the peak of the high-energy component is comparable or lower 
than the synchrotron peak luminositiy \citep{1978bllo.conf..328B}.

If the high-energy component is produced by
an inverse Compton scattering of photons reprocessed in the BLR,
then the spectra of FSRQs should have a cut-off at a few GeV due to the
Klein-Nishina~(KN) effect  \citep{2005MNRAS.363..954M}.  
Spectral breaks at a
few GeV have been found in many
FSRQs \citep{2011ApJ...733L..26A}.  
The most prominent example is
3C~454.3 \citep{2009ApJ...699..817A}.  
In addition, the luminous IR-UV photon fields from the BLR or the DT can cause 
a strong absorption of HE and very high-energy (VHE: $>$ 100 GeV) photons by electron-positron
pair production \citep{2003APh....18..377D,2006ApJ...653.1089L}. 
This mechanism was proposed by \citet{2010ApJ...717L.118P} to
explain the observed spectral features in several blazars.

Despite a large number of FSRQs detected in the HE range, 
almost all blazars observed in the VHE range belong to the BL Lac class of 
objects \citep{2012arXiv1205.0068E}.  Recently, however, Cherenkov
telescopes have detected three FSRQs in the sub-TeV range.  The first
detected object was 3C~279, observed with the MAGIC
telescope \citep{2011A&A...530A...4A}.  Two additional FSRQs are
4C~21.35, detected by the MAGIC telescope \citep{2011ApJ...730L...8A},
and PKS~1510-089, detected with
H.E.S.S.\ \citep{2011ATel.3509....1H,2010HEAD...11.2706W}.  The
detection of these objects proved that FSRQs can also emit photons in
the VHE range.  

This VHE emission is very difficult to  explain within the current models of FSRQs, 
which assume that the $\gamma$-ray emitting region is close 
to the base of the jet, within distances up to 10$^{-2}$~pc, 
where the strong external photon field originating from the broad line region is present. 
We investigate impacts of the location of the emitting region on the spectral energy distribution.
In particular, we focus on the impact of an environment on the jet,
 such as external photon fields, 
the $\gamma - \gamma$ absorption, and the Klein-Nishina effects.

The paper is organized as follows.  In Section \ref{sec:pks1510}, we
describe general properties of PKS~1510-089.  
In Section \ref{sec:obs}, we report the results of the monitoring on the source during its
flaring activity in March 2009, together with simultaneous
multiwavelength observations across the electromagnetic spectrum.  In
Section \ref{sec:model}, we present the SED modeling and discuss the
absorption, as well as the location of the $\gamma$-ray emitting region
in the jet of PKS~1510-089.  The discussion and conclusions are presented in
Sections \ref{sec:con}.


\section{Blazar PKS 1510-089  \label{sec:pks1510}}

Blazar PKS~1510-089 ($\alpha_{\rm J2000} = 15^{\rm h}12^{\rm
m}50.5^{\rm s}$, $\delta_{\rm J2000} = -09^{\rm d}06^{\rm m}00^{\rm
s}$), at a redshift of $z=0.361$, is a FSRQ detected in the MeV-GeV band
by the EGRET detector \citep{1999ApJS..123...79H}.  It is characterized by
a highly relativistic jet that makes a $\sim3^\circ$
angle relative to the line of sight \citep{2005ASPC..340...67W}.  
The radio jet of PKS~1510-089 is curved and shows 
an apparent superluminal motion as high as $45$
times the speed of
light \citep{2001ApJ...549..840H,2002ApJ...580..742H,2005AJ....130.1418J}.

The first large multiwavelength campaign on PKS~1510-089 took place
in August 2006 \citep{2008ApJ...672..787K} and involved the {\it Suzaku}
and {\it Swift} satellites, as well as ground-based optical and radio instruments.  
The campaign allowed the construction of the
broadband spectrum of the object, ranging from $10^{9}$ to
$10^{19}\,$Hz. The spectrum was successfully modeled using the one-zone
internal shock scenario.  \citet{2008ApJ...672..787K} focused their
work on the explanation of the X-ray part of the SED, where an excess
of emissions has been observed. If interpreted as bulk-Compton
radiation, this excess allows an estimate of the pair content of the jet.  In the
case of PKS~1510-089, the ratio of e$^+$e$^-$ pairs to the number of
protons was estimated to be on the order of $10$.  This implies that
although the number of e$^+$e$^-$ pairs is larger than the number of
protons, the power of the jet is still dominated by the latter.  As an
alternative interpretation, \citet{2008ApJ...672..787K} proposed that
the observed soft X-ray excess might be explained as a contribution of
the synchrotron self-Compton (SSC) component, which, although
energetically inefficient, shows its presence in the soft X-ray range.

\citet{2010ApJ...721.1425A} have reported on multiwavelength observations
of PKS~1510-089 during a high activity period between September~2008
and July~2010.  These observations revealed a complex variability in
optical, UV, X-ray, and $\gamma$-ray bands of time scaled down to
$6-12\,$hours.  The study of the correlation of the variability in
different energy ranges, performed by \citet{2010ApJ...721.1425A}, shows lack of
a correlation between the $\gamma$ rays and the X-rays, a weak
correlation between the $\gamma$ rays and the UV(R) band, and a
significant correlation of $\gamma$ rays with the optical
band.  \citet{2010ApJ...721.1425A} attempted to model three distinctive
flares observed during the period with simultaneous data from radio to
$\gamma$-ray energies.  They adopted the IC scenario with seed photons
originating from the BLR to explain the HE emission.  The emission
region in their model was assumed to be located within the subparsec scale.
A significant fraction of the IC scattering in their model occurred
in the KN regime leading to the curved MeV/GeV spectral shape that
matches the observed spectrum in the HE range.

A multiwavelength campaign in a quiescent state of
PKS~1510-089 was conducted in 2011 \citep{2012ApJ...760...69N}.
The campaign  included {\it Herschel} observations combined
with the data publicly available from the {\it Fermi}/LAT, {\it Swift}/XRT,
SMARTS, and the Submillimeter Array (SMA).
 \citet{2012ApJ...760...69N} concluded that at least a two-zone blazar model is necessary to
interpret the entire dataset.  They suggested that the observed
infrared emission is associated with the synchrotron component
produced in the hot-dust region.  To explain the $\gamma$-ray emission,
they proposed an IC component produced in the BLR.  Both components
were located within the subparsec scale.  In such a scenario, the optical/UV
emission would be associated with the accretion disk thermal emission,
with the accretion disk corona likely contributing to the X-ray
emission.
\citet{2012ApJ...760...69N} demonstrated  that a single-zone
scenario would require an unrealistically high-energy density of the
external radiation to explain the observed data and, in particular, the Compton dominance.

Most recently, \citet{2013ApJ...766L..11S} investigated the $\gamma$-ray light curve 
of the source obtained with {\it Fermi}/LAT during 
the period from September to December 2011.  They found a variety of
temporal characteristics and variability patterns, e.g.
very rapid flares with doubling times of 1~h and an energy release comparable 
to the kinetic luminosity of the jet.


\section{Observations \label{sec:obs}}

In March 2009, a flaring activity of PKS~1510-089 was reported in
the high-energy range \citep{2009ATel.1957....1D,2009ATel.1968....1P,2009ATel.1976....1V}.
The flaring activity was also detected by the ATOM telescope and
the GASP project \citep{2009ATel.1988....1V}.  Observations with the
H.E.S.S.\ telescopes followed the reporting of this flaring activities.
In this work we used the H.E.S.S. and Fermi/LAT spectrum published by H.E.S.S. Collaboration et al. (2013), which covers two periods. 
The majority of VHE emissions was detected during the first period. 
The {\it Fermi}/LAT spectrum is obtained for the time period simultaneous with the H.E.S.S. observations of PKS 1510-089.

\subsection{H.E.S.S. observation}
The H.E.S.S.\ data were taken simultaneously with the peak of the HE
flare recorded by {\it Fermi}/LAT.  The H.E.S.S. Collaboration
carried out observations of PKS~1510-089 in two periods.  The first
observations were taken between March 23, 2009 (MJD 54910) and April
2, 2009 (MJD 54923).  The second report of the HE activity
triggered the H.E.S.S.\ observations between April 27, 2009 (MJD
54948), and April 29, 2009 (MJD 54950).  The H.E.S.S.\ observations
resulted in $15.8$~hours of good-quality data. 
The spectrum in the energy range from 0.15~TeV to 1~TeV 
is well represented by the power-law function with 
a spectral index of $\Gamma = 5.4\pm 0.7_{\mbox{stat}} \pm 0.3_{\mbox{sys}}$.
The integral flux corresponds to $\approx$3\% of the Crab Nebula, 
or, equivalently 
$I(0.15 TeV < E < 1.0 TeV)= (1.0 \pm 0.2_{\mbox{stat}} \pm 0.2_{\mbox{sys}}) \times 10^{-11} \mbox{cm}^{-2}\mbox{s}^{-1}$ \citep{2013arXiv1304.8071H}. 

\subsection{{\it Fermi}/LAT data analysis}
\label{sec:FERMIdata}
The {\it Fermi}/LAT \citep{2009Atwood} data, simultaneously with the
H.E.S.S.\ observations, were published by the \cite{2013arXiv1304.8071H}.
The light curve between MJD~54909.5 and MJD~54951.5 shows
two evident flares,
one centered around MJD~54916, and the other centered around MJD~54948.

The HE data taken between MJD~54914.8 and MJD~54917.5
were best fitted by the log-parabola model $dN/dE=n_0 (E/E_0)^{-\alpha - \beta ln(E/E_0)}$, 
with the following parameters: a normalization $n_0=(10.5\pm 0.7)\times10^{-9}$cm$^{-2}$s$^{-1}$MeV$^{-1}$,
slope parameters of $\alpha=1.81\pm0.13$ and $\beta=0.161\pm0.048$, 
and the energy $E_0$ fixed at 260~MeV. 

\subsection{RXTE, {\it Swift}, optical and radio data}
The RXTE light curve and spectrum, as well as the radio data, were presented in \cite{2010arXiv1002.0806M}.

The {\it Swift}/XRT spectrum in the period corresponding to the H.E.S.S. observations (MJD~54910-54923)  was
produced using the Build {\it Swift}/XRT Products tool \citep{2009MNRAS.397.1177E}.  
The resulting {\it Swift}/XRT spectrum is accurately 
described by a power law with photon index $1.45\pm0.03$. 
The galactic absorption  with a hydrogen  column density of $6.89\times10^{20} \mathrm{cm}^{-2}$ was assumed \citep{2005A&A...440..775K}.

The optical observations have been carried out with the ATOM telescope
located on the H.E.S.S. site and operated by the
H.E.S.S. Collaboration.  PKS 1510-089 is one of the sources regularly
observed with the ATOM.  The maximum optical flux of $5.724\pm 0.150$mJy was observed by the ATOM in
the R band in March 2009 \citep{2013arXiv1304.8071H}.

The radio data during the flare were recorded through the Michigan Radio Astronomy
Observatory ($14.5\,$GHz), the Metsaovi Radio Observatory ($37\,$GHz),
and the Submillimeter Array ($230\,$GHz) and are presented in \cite{2010arXiv1002.0806M}.

\section{Modeling \label{sec:model}}

We have aimed to reproduce the spectral energy distribution (SED) of PKS~1510-089 during the high-energy (HE) 
state recorded in March 2009, during which very high-energy emission (VHE) were also detected by H.E.S.S. telescopes. 
Our approach was to use the one-zone leptonic model implemented in the \texttt{BLAZAR}
code \citep{2003A&A...406..855M}.

The \texttt{BLAZAR} code models non-thermal flares 
following the evolution of relativistic electrons injected into the
conical jet as a result of shock operation.  The shock forms during
the collision of inhomogenities propagating down the jet with
different velocities.  Such inhomogenities may be created by some
instabilities in the very central part of the active galactic nuclei.
In the absence of a detailed model of particle acceleration the
injected distribution of particles is assumed to be a broken power
law.  The injected particles then lose their energy because of
synchrotron and inver-Compton (IC) emission as well as adiabatic cooling.
 
The \texttt{BLAZAR} code requires some input parameters, notably the
value of the energy density of an external diffuse radiation field,
the injected electron energy distribution, the value of the magnetic
field, and the description of the overall geometry of the emitting
region.

The parameters of these external photon fields, used as the input, 
are estimated based on the observations and known relations (see section~\ref{sec:ModelParameters}).
The choice of  geometry, mainly opening and viewing angles, and the Doppler factor 
are in first approximation motivated by observations,
and then are tuned to best reproduce the overall spectral energy distribution.
The localization of the emitting region is obtained by combining various effects,  
thus our choice is strongly constrained by observations (see  sections~\ref{sec:InternalAbsorption} and~\ref{sec:location}).
The value of the magnetic field and the electron distribution cannot be constrained directly,
but can be found from the modeling of the SED. 
The modeling of simultaneous multiwavelength data, as in the case of the 2009 active state of PKS~1510-089,
provides the best tool for constraining  the values of the magnetic field and the electron energy distribution.

The next sections present the model parameters and introduce two
effects that are important for modeling of FSRQs at high-energies, 
namely absorption of high-energy photons in a process of pair production
and the Klein-Nishina effect.  Later in this section we
discuss the location of the region where the bulk of the HE emission is
produced, the so-called blazar zone, and present the obtained SED of
the object.

\subsection{Model parameters}
\label{sec:ModelParameters}

The \texttt{BLAZAR} code calculates the evolution of electrons
injected along the jet.  The injected electrons follow a broken power-law distribution,
\begin{equation}
  Q(\gamma) = K_e \left\{
    \begin{array}{l l}
      \gamma_b^{p-q} \gamma^{-p}
      & \mathrm{for} {~~~} \gamma_\mathrm{min} \le \gamma \le \gamma_b \\
      \gamma^{-q} & \mathrm{for} {~~~} \gamma_b < \gamma \le \gamma_\mathrm{max}
    \end{array}
    \right. \,,
\end{equation}
where $K_e$ is the normalization of the injection function,
$p$ and $q$ are low- and high-energy electron spectral indices,
and $\gamma_b$, $\gamma_{\mathrm{min}}$, and $\gamma_{\mathrm{max}}$ are
break, minimum, and maximum electron Lorentz factors, respectively.

The electron injection starts at some distance  $R_0$ from the center and continues until $R = 2
R_0$, while the electron evolution is followed up to $3 R_0$.  The
proper choice of $R_0$ is crucial for the SED modeling.
The luminosity of the accretion disk has been reported by
\citet{2012ApJ...760...69N} to be $L_{\mathrm{d}} = 5 \times
10^{45} \,\mathrm{erg\,s}^{-1}$.

Two sources of external photons, DT and BLR, were considered.  
Following the equation \citep{2005MNRAS.361..919P}
\begin{equation}
  R_{\mathrm{BLR}}=(22.4\pm0.8) \left [ \frac{\lambda
      L_{\lambda}(1350\mbox{\AA})}{10^{44}\,\mbox{erg\,s}^{-1}} \right
  ]^{0.61\pm0.02} \, \mathrm{light \,days}\,,
  \label{eq:RBLR}
\end{equation}
the size of the BLR, $R_{\mathrm{BLR}}$, is estimated to be
$0.12 \times 10^{18}\,$cm.  The dust temperature is $T = 1.8 \, T_3$,
where $T_3 = 1000\,$K \citep{2012ApJ...760...69N}.  The size of the DT is
approximately  $R_{DT} \simeq 1.94 \times 10^{18}\,$cm (see Eq.~39 in \citet{2009ApJ...704...38S}).

The energy density of the magnetic field, $B$, is defined as $u_B=B^2/8\pi$ (see Table~\ref{tab:modelfit}).
The energy density of external radiation fields have been calculated
using the equation \citep{2009ApJ...704...38S}
\begin{equation}
  u_{\mathrm{ext}} (r) \simeq \frac{\xi_{\mathrm{ext}}L_{\mathrm{d}}}{4 \pi c
    R_{\mathrm{ext}}^2}\frac{1}{1+(r/R_{\mathrm{ext}})^{n_{ext}}} \,,
  \label{eq:uext}
\end{equation}
where "$\mathrm{ext}$" can be either DT or BLR, and $\xi_\mathrm{ext}$
is a fraction of reprocessed emission from the accretion disk.  
We adopted $n_{BLR}=3$  and $n_{DT}=4$  \citep{2009ApJ...704...38S,2012ApJ...760...69N,2012ApJ...754..114H}.
The corresponding energy density in the jet comoving frame
is approximated as $u'=4/3\Gamma^2 u_{\mathrm{ext}}$.

\begin{figure}
  \centering      
  \includegraphics[height=\linewidth,angle=-90]{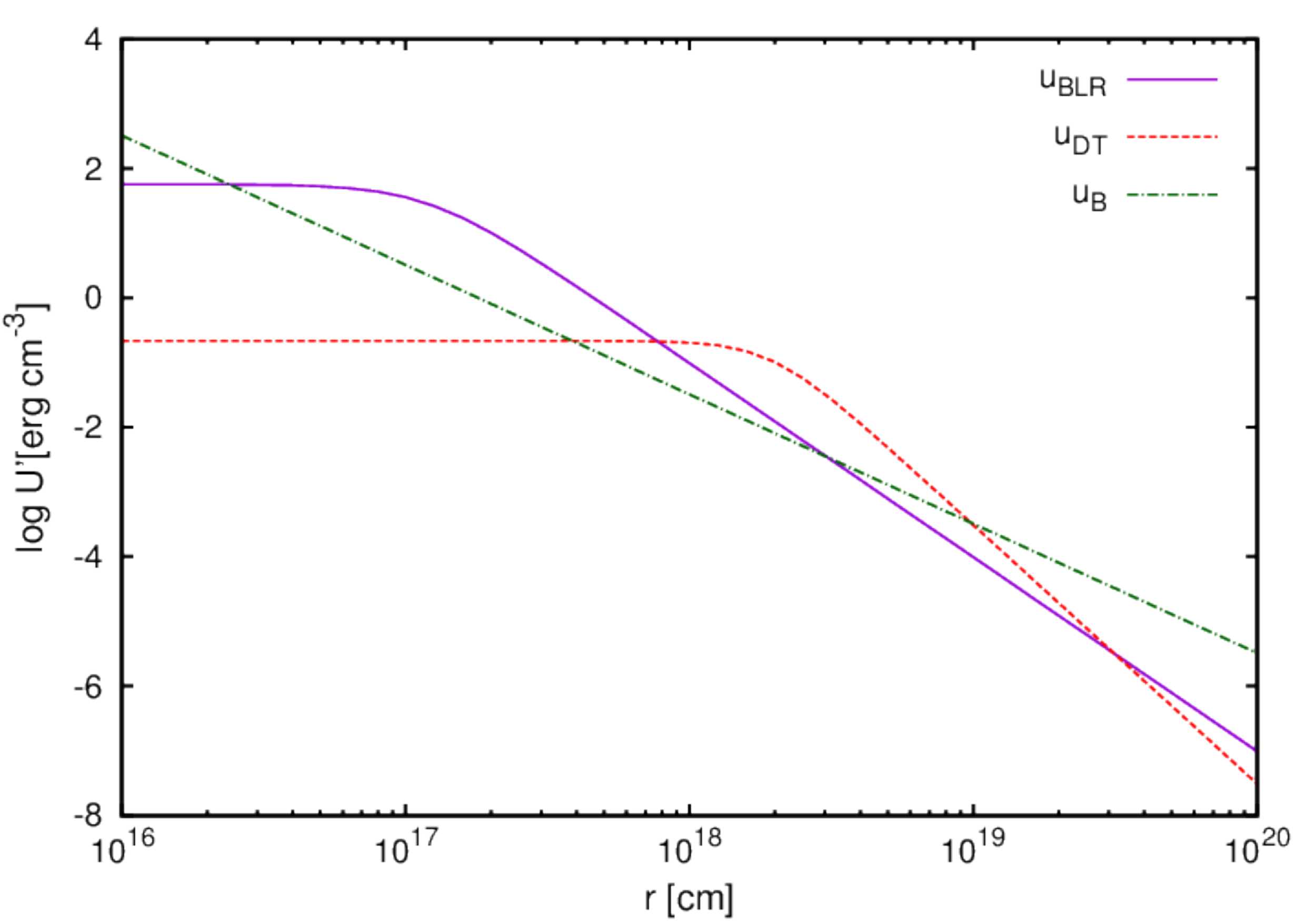}
  \caption{The energy density of the extended fields as a function of the distance from the
    central source.  
    The energy density is given in the  jet comoving frame.
    The violet solid line represents the energy
    density of BLR, the red dashed line  the energy density of
    DT, and the green dashed-dotted line  the energy density of
    magnetic field. }
  \label{fig:energy_density}
\end{figure}

We have used $\xi_{\mathrm{BLR}} \approx 0.1$ and $\xi_{\mathrm{DT}} \approx
0.2$.  The values of $\xi_{\mathrm{BLR,DT}}$ together with values of
$L_{\mathrm{d}}$, $R_{\mathrm{BLR,DT}}$ listed above, give the density of
external radiation fields at a distance $2 R_0$, $u_{\mathrm{BLR}} \approx
0.09\,\mathrm{erg} \, \mathrm{cm}^{-3}$ and $u_{\mathrm{DT}} \approx
0.0005\,\mathrm{erg} \,\mathrm{cm}^{-3}$, 
where $R_0$ is the distance from the central engine to the point
where the electron injection starts (see Table~\ref{tab:modelfit}).
The distribution of the energy density of the photon fields as a function of distance is
presented in Fig.~\ref{fig:energy_density}.

We assumed a jet Lorentz factor of $\Gamma=22$ and with the jet opening angle, $\theta_{\rm jet}$, related as $1/\Gamma$. 
The jet opening angle is then $\approx 2.6^\circ$ (see Table~\ref{tab:modelfit}).
The object is observed at $\theta_{\mathrm{obs}}=\theta_{\mathrm{jet}}$, 
which is consistent with observations \citep{2005ASPC..340...67W}. 

\subsection{Absorption}
The fundamental process responsible for the absorption of HE $\gamma$ rays 
is the electron-positron pair production.

The observed HE spectrum after attenuation is 
\begin{equation}
F_{\mathrm{obs}}(E)= F_{\mathrm{int}}(E) e^{-\tau(E)} \,,
\end{equation}
where $e^{-\tau(E)}$ is the attenuation, $\tau(E)$ is the optical
depth, and $F_{\mathrm{int}}$ is the intrinsic spectrum of the source.
The optical depth given by \cite{1967PhRv..155.1404G} is
\begin{eqnarray}
\tau(E)& = & \int \mbox{d}l \int_{\cos\theta_{min}}^{\cos\theta_{max}}
\mbox{d}\cos{\theta} \frac{1-\cos{\theta}}{2} \times \nonumber \\
& & \times \int_{E_{th}}^{\infty}
\mbox{d}\epsilon n(\epsilon)\sigma(E,\epsilon,\cos{\theta}) \,,  
\label{eq:taue}
\end{eqnarray}
where $\mathrm{d}l$ is the differential path traveled by the HE
photon, $\theta$ is the angle between the momenta of HE and low-energy
(LE) photons, and $E_{th}$ is a threshold energy for pair production.  
The energy of the HE photon is $E$ and the energy of the LE
photon is $\epsilon$.  The density number of the LE photons  is
$n(\epsilon) \, \mbox{cm}^{-3}$.  The cross-section
$\sigma(E,\epsilon,\cos{\theta})$ of pair production is given by
\citet{2008arXiv0809.5124B}.

High-energy photons can be absorbed by several photon fields during their
travel from the emission region to the observer.  The first sources of
LE photons are located in the blazar itself (internal absorption).
Then, when photons escape from the blazar, they travel over
cosmological distances and may be absorbed by LE photons of the
extragalactic background light (external absorption).

\subsubsection{Internal absorption}
\label{sec:InternalAbsorption}
The first possibility of absorption arises from the photon fields
present in the blazar itself.  When a HE photon is produced in the
jet within the BLR radius, it has to travel through the combined photon fields of BLR and DT.
The characteristic frequency of the BLR radiation is of the order of
$10\,$eV \citep{2008ApJ...672..787K}, while the characteristic frequency of DT is of the order of
$0.1\,$eV \citep{2012ApJ...760...69N}.  Since the energy of BLR photons is larger than that of DT
photons, the threshold energy of pair production is smaller for the
BLR and HE photons are absorbed above energies of a few GeV. 
Specifically, at a distance $R=10^{17}\,$cm (just inside $R_{\mathrm{BLR}}$) 
the attenuation for 200 GeV photons due to internal absorption is $\approx 0.1$.

 To avoid significant absorption by BLR photons, the blazar zone, where HE
photons are emitted, has to be located outside the BLR.  The photon
number density of external radiation (from both BLR and DT) decreases
with distance as in Eq.~(\ref{eq:uext}).  When HE photons propagate
inside the BLR or DT region then the distribution of the photon field
is isotropic and therefore the $\cos\theta$ in  Eq.~(\ref{eq:taue})
ranges from $-1$ to $1$.  However, outside of the BLR or DT regions,
$\theta$ have values ranging from $\theta_\mathrm{min} = \pi +
\arctan(R_{\mbox{BLR,DT}}/r)$ to $\theta_\mathrm{max} =
\pi-\arctan(R_{\mbox{BLR,DT}}/r)$.

\begin{figure*}
  \centering
  \begin{minipage}{.5\textwidth}
    \centering
    \includegraphics[height=1\linewidth,angle=-90.0,clip]{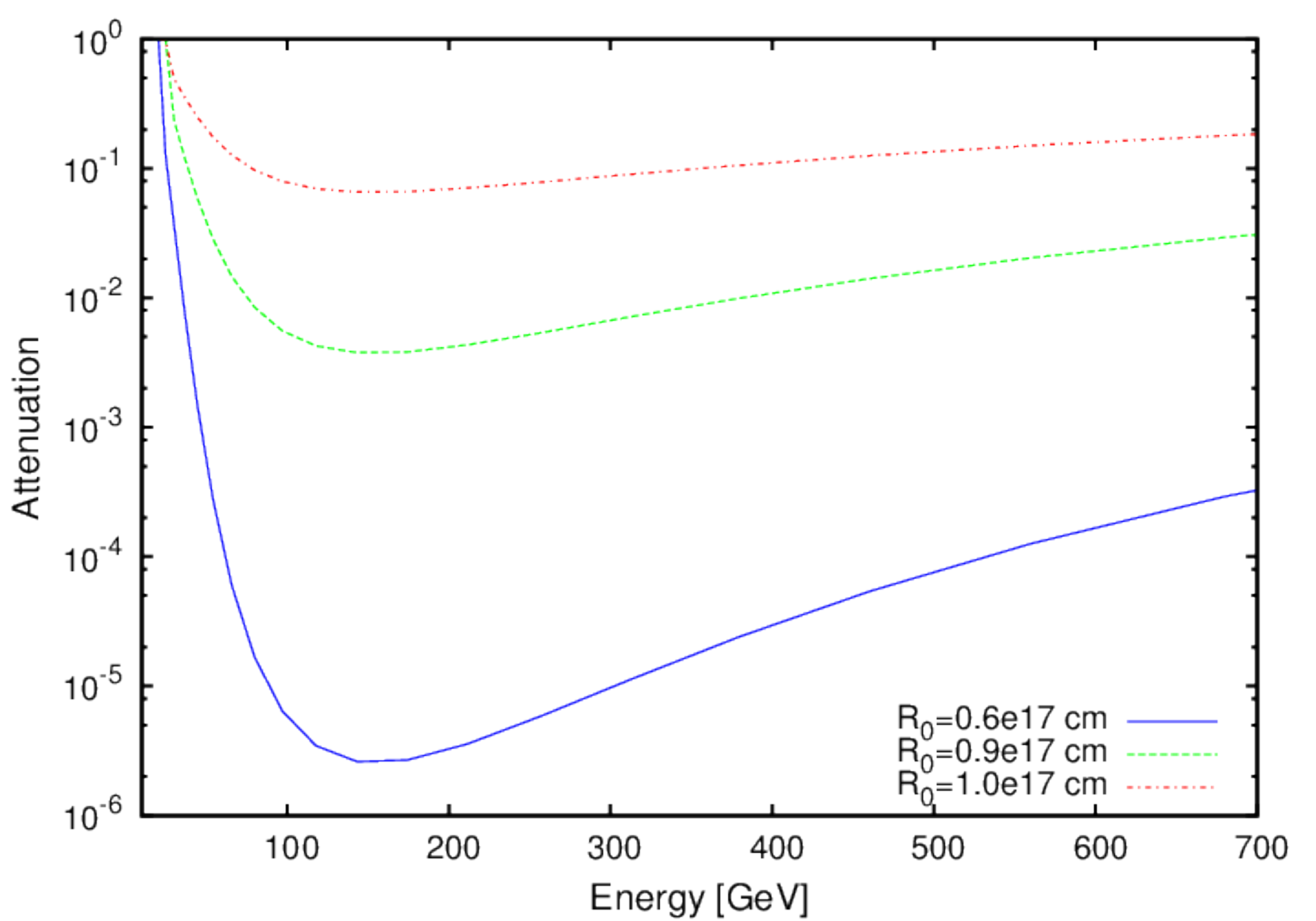}
  \end{minipage}%
  \begin{minipage}{.5\textwidth}
    \centering
    \includegraphics[height=1\linewidth,angle=-90.0,clip]{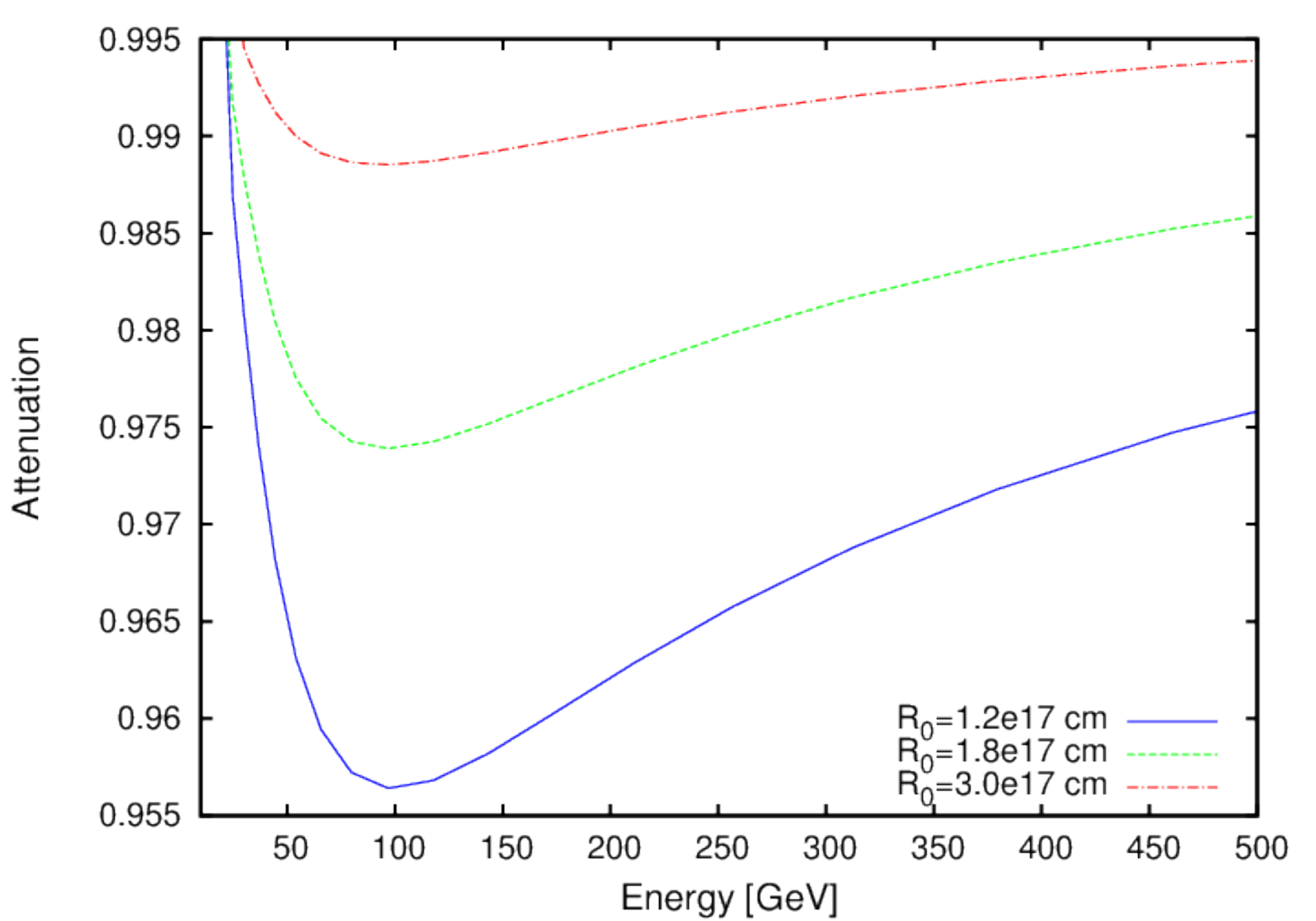}
  \end{minipage}
  \caption{The internal absorption as a function of photon energy
    emitted in the blazar zone.  The values of attenuation presented for BLR,
    $e^{-\tau(E)}$, are calculated for different distances from the
    center and assuming the PKS~1510-089 model parameters listed in
    Table~\ref{tab:modelfit}.  Left panel: $R_0$ below
    $R_{\mbox{BLR}}$.  Right panel: $R_0$ above
    $R_{\mbox{BLR}}$.}
  \label{fig:attenuation}
\end{figure*}

Figure~\ref{fig:attenuation} shows the internal absorption caused by photons from BLR,
$e^{-\tau(E)}$, as a function of the photon energy.

The internal absorption caused by 
photons from the DT is significant for $\gamma$-ray photons with 
energies above 400~GeV and
can thus  be ignored in further investigation
because the emissions detected during the 2009 flare by H.E.S.S. is in sub-TeV range. 

\subsubsection{External absorption}
The other source of LE photons is the extragalactic background light
(EBL).  The EBL is the IR/UV radiation generated by stars (UV) and
radiation emitted through the absorption and re-emission of star light
by dust in galaxies (IR).  The EBL models have been reviewed recently
by e.g.\ \citet{2001ARA&A..39..249H} and new constraints on the EBL
intensity have been provided by \citet{2012A&A...542A..59M} using the
{\it Fermi}/LAT data and \citet{2012arXiv1212.3409H} using the
H.E.S.S. observations of the brightest blazars.

In the case of EBL, the energy of HE photon, $E$, as well as the
energy of LE photon, $\epsilon$, have to be multiplied by $(1+z)$,
where $E$ and $\epsilon$ are the observed photon energies at $z = 0$.
The distribution of angles, $\theta$, at which background photons can
collide with HE photons is flat when a photon is traveling over
cosmological distances; therefore, the $\cos\theta$ of the scattering
angle (see Eq.~\ref{eq:taue}) changes from $-1$ to $1$.

The EBL model of \cite{2008A&A...487..837F} has been used in our modeling of PKS~1510-089.

\subsection{Location of the $\gamma$-ray emitting region in PKS 1510-089}
\label{sec:location}
The choice of the distance of the shock formation from the central
source, $R_0$, is constrained on the one hand by the internal absorption
of gamma rays in BLR, and on the other hand by the IC efficiency.
  When $R_0$ is smaller than $R_{\mathrm{BLR}}$,  a
significant fraction of the HE radiation (from several dozen to
several hundred GeV) is absorbed.  If $R_0$ is greater than
$R_{\mathrm{BLR}}$, then only a few percent of the HE emission is
absorbed.

If, however, the blazar zone is too far away from the BLR
then the photon energy density is too small to produce the observed
$\gamma$-ray emission by IC mechanism.

There is some observational evidence that the blazar zone in
PKS~1510-089 may be located outside the BLR.  The radio observations
of PKS~1510-089 between  September 9, 2011, and  October 17, 2011,
 show a $\Delta t_{\mathrm{obs}} \sim
40\,$day increase in radio flux \citep{2012arXiv1210.4319O}. 
If the $\gamma$-ray flare is 
associated with the same region as the radio flare, the projected
distance between the regions where the shock formed
and the site responsible for the $\gamma$-ray emission is $\sim$0.6~pc  \citep{2012arXiv1210.4319O}.   
Assuming a jet inclination
angle of $\theta_{\mathrm{jet}} \sim 3^\circ$
\citep{2010arXiv1002.0806M}, the deprojected
distance is about $\simeq 10\,$pc.  This demonstrates that at least for some
$\gamma$-ray flares, the blazar zone may be located far outside the
BLR.

In the presented modeling of PKS~1510-089, we adopted $R_0 = 0.7
\times 10^{18}\,$cm.  At that distance, the absorption by the low-energy photons originating from the BLR is very small,
less than $1\%$, while the energy density (see Fig.~\ref{fig:energy_density})
in the blazar zone of PKS~1510-089 ($0.7-1.4 \times 10^{18}\,$cm) is
still dominated by radiation from the BLR.  Outside
$R_{\mathrm{BLR}}$, the external radiation field is dominated by
$u_{\mathrm{BLR}}$ up to a distance $R_{\mathrm{DT}}$, where $u_\mathrm{BLR}$
becomes comparable to the energy density of the radiation from DT.

In the considered scenario, the blazar zone is placed outside the BLR, thus 
the radiation form BLR comes from behind the outflowing plasma jet.
The IC energy losses are smaller here than in the isotropic external radiation fields \citep{2009MNRAS.397..985G,2013ApJ...779...68S}. 
The impact of the direction of the radiation field on the observed 
spectral energy distribution is currently being investigated (Moderski \& Bhatta, in preparation).

\subsection{Klein-Nishina effect}
When the product of the energy of the photon and the energy of the
electron before the collision $\gamma \epsilon$ is $\ll 1$, where
$\gamma$ is the electron Lorentz factor and $\epsilon=h\nu/(m_ec^2)$, the
interaction proceeds in the so-called Thomson regime.  In this
regime, the rate of IC energy losses of relativistic, isotropically
distributed electrons is
\begin{equation}
|\dot \gamma_T| = \frac{4 c \sigma_T}{3 m_e c^2} \gamma^2 u_0, 
\label{eq:ICcool}
\end{equation}
where $u_0$ is the total energy density of the radiation field.
 
However, in the case of HE emission in blazars, the photon energy may
become comparable or even larger than the electron energy ($\gamma
\epsilon > 1$).  In such a case the cross-section has to be
expressed using the full Klein-Nishina (KN) formula
\citep{2005MNRAS.363..954M}.  The cross-section is smaller here than in
the Thomson regime as the photon energy becomes larger.  One of the
consequences is a reduction of the electron energy loss rate
\begin{equation}
\dot \gamma=\dot \gamma_T \, F_{KN} \,,
\label{eq:ICcoolKN}
\end{equation}
where $\dot \gamma_T$ is given by Eq.~(\ref{eq:ICcool}) and a
factor $F_{KN}$ is given by \citet{2005MNRAS.363..954M}.

\begin{figure}
  \centering
  \includegraphics[height=1\linewidth,angle=-90.0]{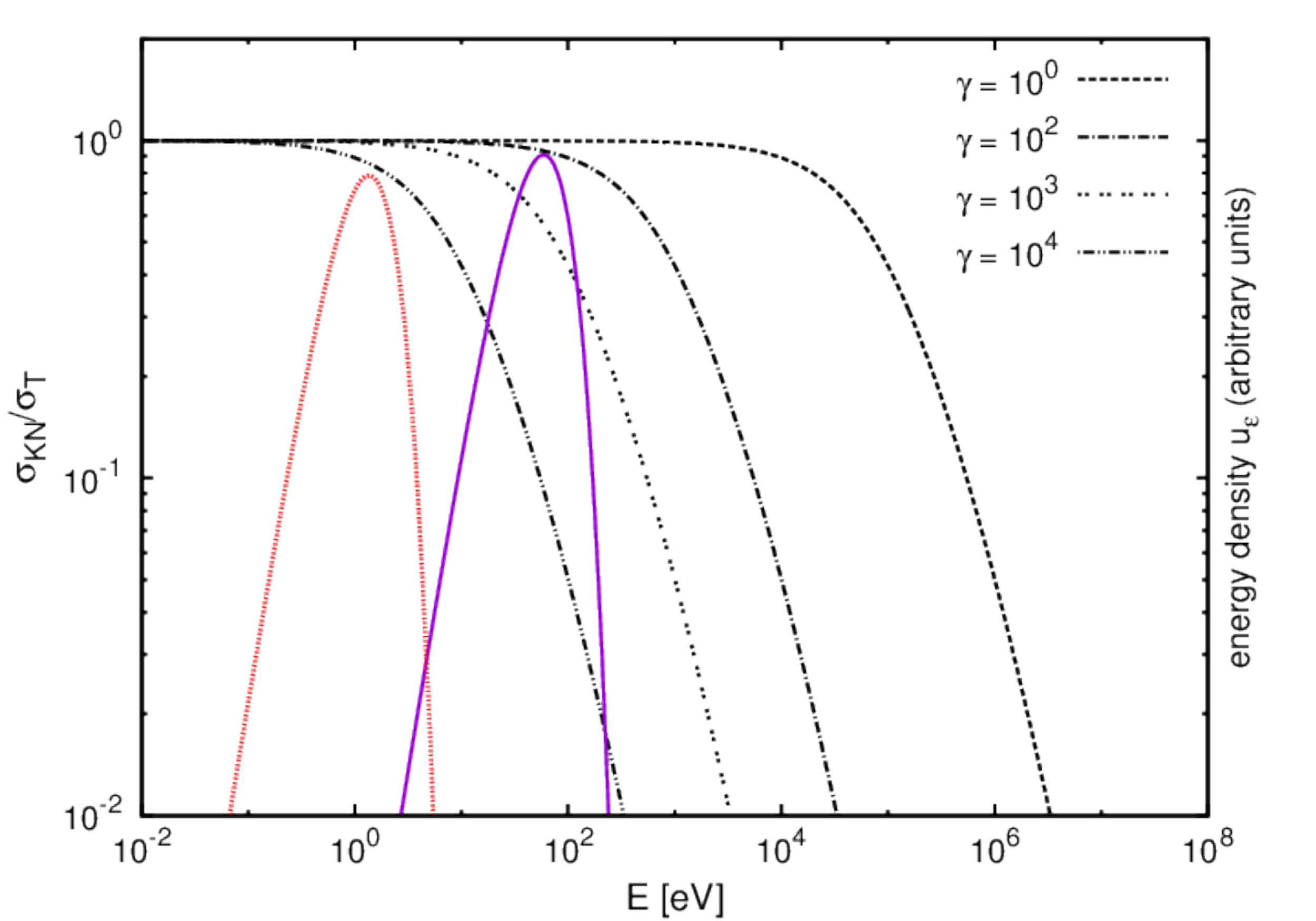}
  \caption{The Klein-Nishina correction factor,
    $\sigma_\mathrm{KN}/\sigma_\mathrm{T}$, as a function of the
    photon energy for different electron Lorentz factors, $\gamma$
    (black lines).  The blue and red lines represent the
    energy densities of two external photon fields, BLR and DT,
    respectively, transformed into the jet comoving frame.}
  \label{fig:fKN}
\end{figure}

Figure~\ref{fig:fKN} presents the Klein-Nishina (KN) correction (a
ratio of the Klein-Nishina cross-section to the Thomson cross-secton
$\sigma_{KN}/\sigma_T$) as a function of the external photon energy
for different electron Lorentz factors $\gamma$.  The KN correction is
presented together with the radiation of the BLR and the DT
approximated as blackbody radiation and transformed into the jet
comoving frame.  Since the electrons with Lorentz factors above $10^3$
are responsible for the HE and VHE emissions it can be seen from the
picture that most of the scattering of the BLR photons happens in the
KN regime, while electrons with Lorentz factors up to $10^4$ scatter
DT photons while they are still in the Thompson regime.

\begin{figure}
  \centering
  \includegraphics[height=\linewidth,angle=-90]{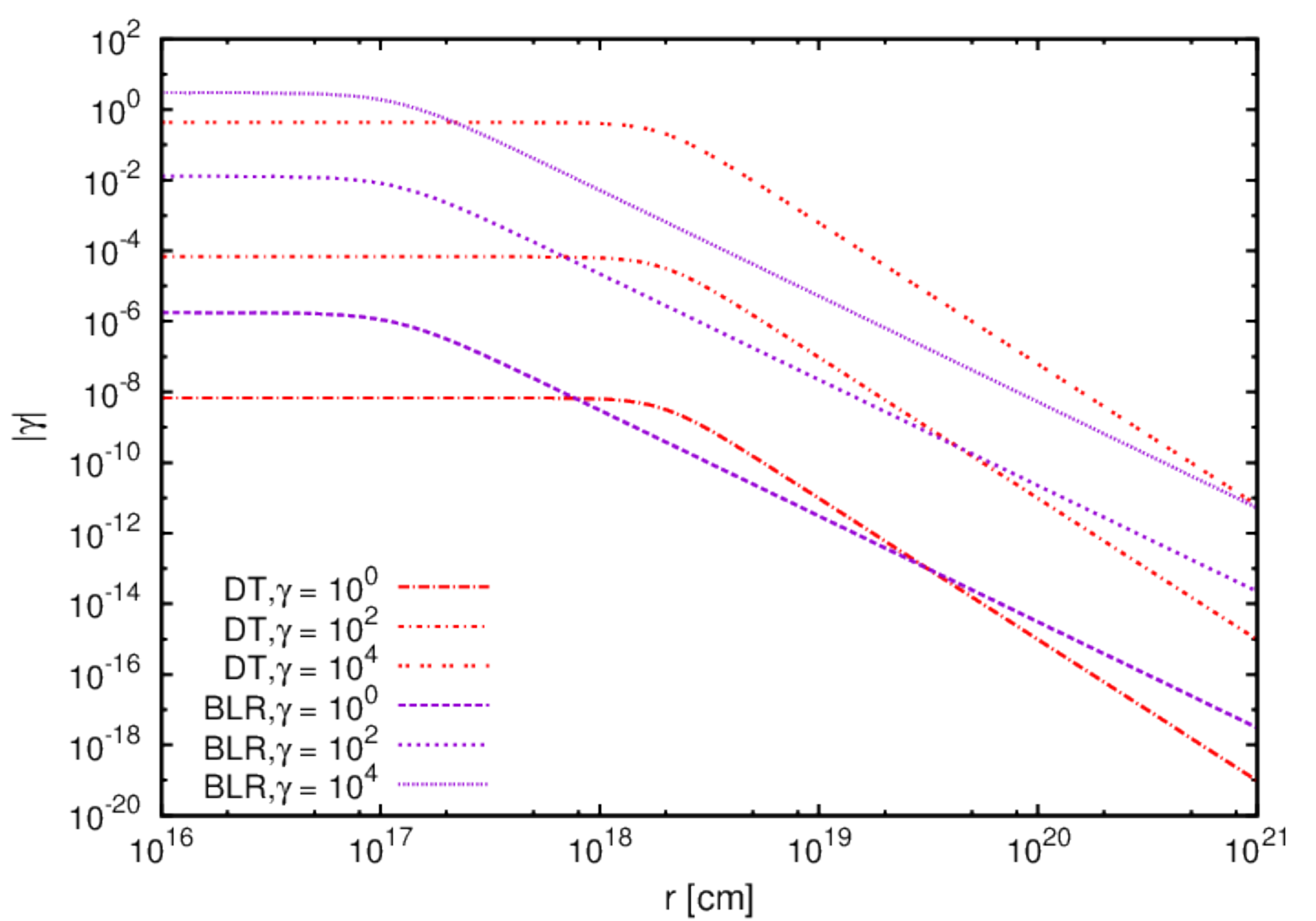}
  \caption{The rate of IC energy losses of relativistic electrons.
    The IC cooling is presented for two sources of seed photons: BLR
    (violet lines) and DT (red lines).  The cooling rate was
    calculated for different electron Lorentz factors $\gamma$: $1$, 
    $100$, $10000$.}
\label{fig:electron_cooling}
\end{figure}

The rate of IC energy losses of relativistic electrons, calculated
according to Eq.~(\ref{eq:ICcoolKN}), is shown in
Fig.~\ref{fig:electron_cooling}.  For sufficiently large distances
from the center ($> 10^{18}\,$cm), high-energy electrons cool
predominantly by the inverse Compton scattering of DT photons, even if
the energy density of the BLR radiation field is comparable or higher
than the energy density of the infrared photons (see
Fig.~\ref{fig:energy_density}).

\subsection{Spectral energy distribution}
\begin{figure*}
  \centering
  \includegraphics[width=\textwidth,angle=0]{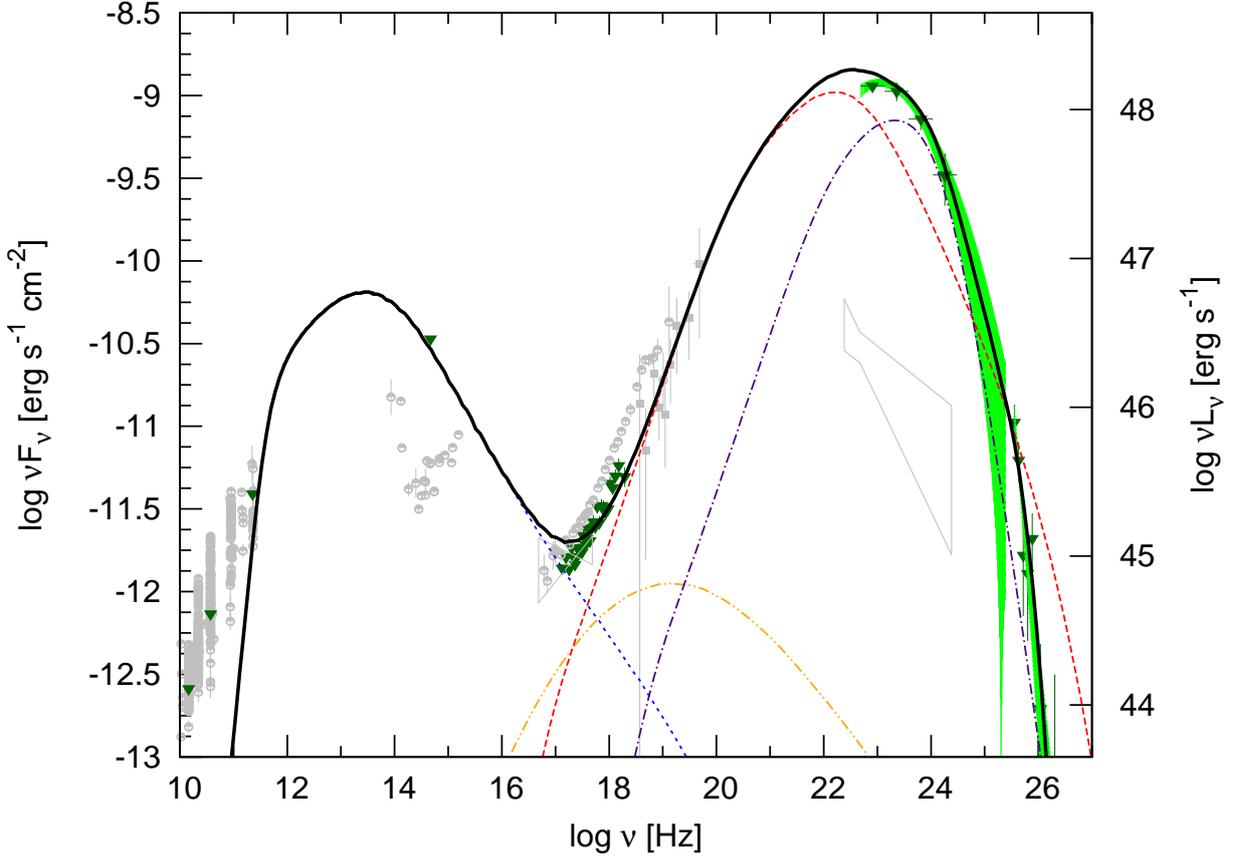}
  \caption{The overall spectrum of  PKS 1510-089.  The short-dashed
    blue line represents the synchrotron component, the dashed red
    line is the inver Compton (IC) component with seed photons originating from the dusty torus (DT) 
    before absorption, and the dashed-dotted violet line is the IC
    radiation with seed photons originating from the broadband region (BLR).  
    The orange dash-double-dotted line
    is the SSC component.  The black solid line represents the sum of
    all the components, corrected for EBL absorption
    (see text for details) and absorption on the low-energy photons originating from the BLR and the DT.}
\label{fig:sed}
\end{figure*}

The other model parameters are estimated to best reproduce
the observed multiwavelength spectrum of PKS~1510-089.  The
parameters are summarized in Table~\ref{tab:modelfit}.  The overall
spectrum of PKS 1510-089 is presented in Fig.~\ref{fig:sed}.  
The green triangles in Fig.~\ref{fig:sed} represent the observations taken around the
VHE flare in March 2009.
The grey circles shows the data published
by \cite{2008ApJ...672..787K}.  The  grey squares  are {\em
  INTEGRAL} data analyzed by \citet{ABIntegral}. 

All discussed  arguments suggest the following scenario for the
PKS~1510-089 flaring activity recorded in March 2009:
\begin{itemize}
\item The low-energy component is produced by the synchrotron
  radiation.
\item The high-energy part of the spectrum (from X-rays to VHE)
  consists of two components:
\item The first component is the IC radiation with seed photons
  originating from the BLR.  This component dominates the emission in
  the {\it Fermi} range.  Because of the KN effect, this component alone
  cannot explain the highest part of the spectrum ($>100\,$GeV).
\item The VHE emission is produced via IC scattering of the seed
  photons originating from DT.  The same component is also responsible
  for the X-ray part of the spectrum, as in the previous modeling
  attempts of this object \citep{2008ApJ...672..787K}.
\item The modeled emission of PKS 1510-089 convolved with the EBL
  attenuation accurately fits the VHE emission.
\end{itemize}

\begin{table*}
\caption{The input parameters for modeling of the non-thermal emission
of the PKS~1510-089.}
\label{tab:modelfit}
\centering
\begin{tabular}{lc}
\hline\hline
Parameter & Model  \\
\hline
minimum electron Lorentz factor $\gamma_{\rm min}$ & $1$ \\
break electron Lorentz factor $\gamma_{\rm br}$ & $900$ \\
maximum electron Lorentz factor $\gamma_{\rm max}$ & $10^5$ \\
low-energy electron spectral index $p$ & $1.2$ \\
high-energy electron spectral index $q$ & $3.4$ \\
normalization of the injection function $K_{e}$ & $1.85 \times 10^{46}\,$s$^{-1}$ \\
bulk Lorentz factor of the emitting plasma $\Gamma$ & $22$ \\
jet opening angle $\theta_{\rm jet}$ & $0.045\,$rad \\
jet viewing angle $\theta_{\rm obs}$ & $0.045\,$rad \\
location of the blazar zone $R_{\rm 0}$ & $0.7 \times10^{18}\,$cm \\
jet magnetic field intensity $B$ & $0.75\,$G \\
scale of the BLR external photon field $r_{\rm BLR}$ & $0.12 \times 10^{18}\,$cm \\
energy density of the external photon field $u_{\rm BLR}$ & $0.09\,{\rm erg\,cm^{-3}}$ \\
photon energy of the external photon field $h \nu_{\rm BLR}$ & $10$\,eV \\
scale of the DT external photon field $r_{\rm DT}$ & $1.94 \times 10^{18}$\,cm \\
energy density of the external photon field $u_{\rm DT}$ & $0.0005 \,{\rm erg\,cm^{-3}}$ \\
photon energy of the external photon field $h \nu_{\rm DT}$ & $0.15$\,eV \\
\hline
\end{tabular}
\end{table*}

\begin{figure}
  \centering
  \includegraphics[height=\linewidth,angle=-90]{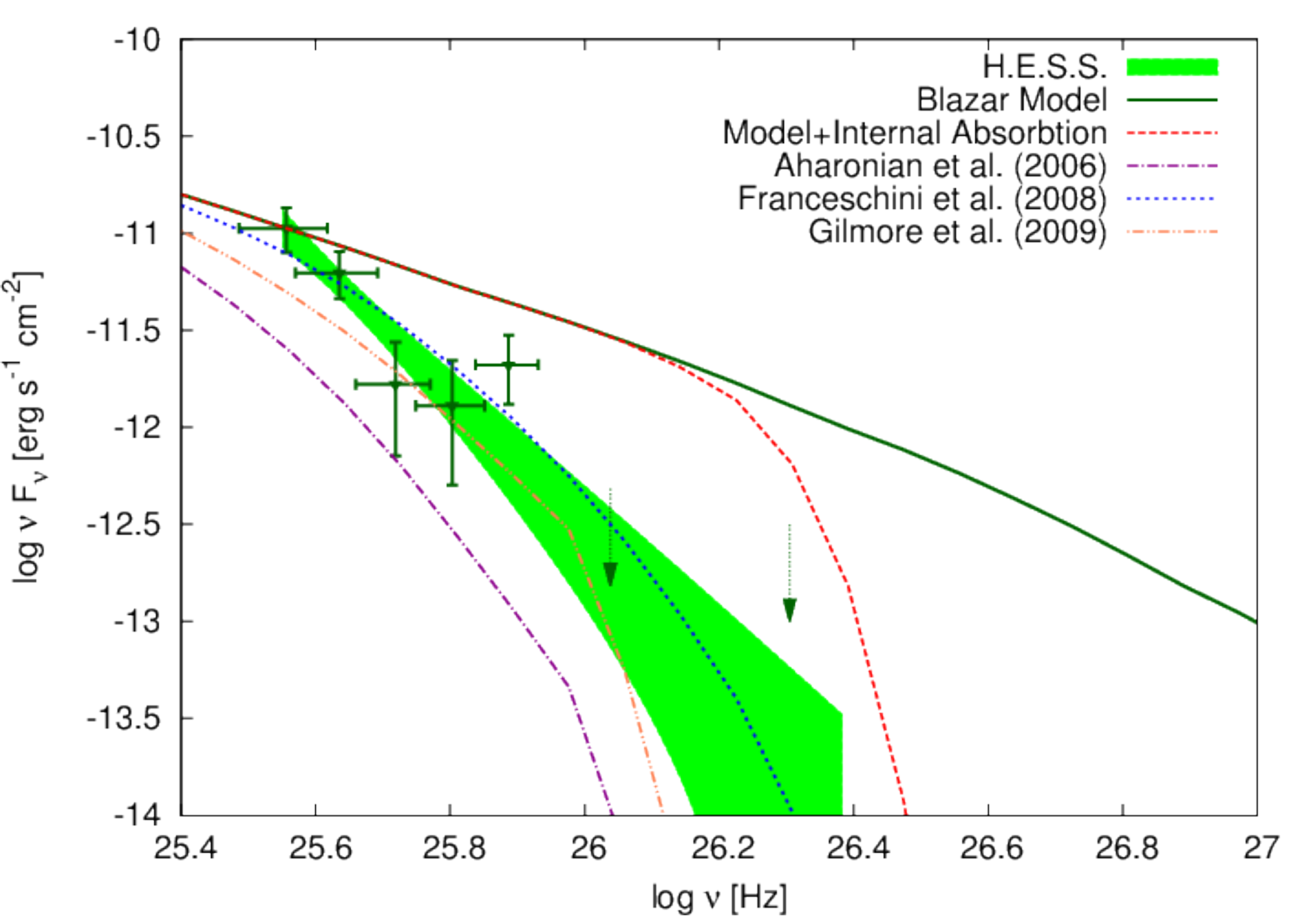}
  \caption{H.E.S.S. observation of  PKS 1510-089 and the model with and without absorption corrections.
	   The shaded region shows the fit to the H.E.S.S. data at  68\% confidence, 
           and the points with errorbars (1$\sigma$ statistical errors) are the energy flux. 
           Arrows denote the 99\% C.L. upper limits. Data taken from \cite{2013arXiv1304.8071H}.
	   The solid line represents the blazar model without internal and external absorption corrections.
	   The red long-dashed line represents the blazar model after internal absorption correction.
          In the proposed model, where the blazar zone is located at a distance of $\sim$1~pc, absorption in the BLR can be ignored.
          The absorption in the DT became significant at energies $\sim400$~GeV, 
           but the emissions, above these energies were not detected by H.E.S.S. in March 2009.
           The short-dashed blue line, the dashed-dotted-dotted orange line, and the dashed-dotted violet 
           line represent the blazar model after internal absorption and external absorption using models provided by
	    \cite{2008A&A...487..837F},  \cite{2009MNRAS.399.1694G}, and \cite{2006Natur.440.1018A}, respectively.}
\label{fig:VHEsed}
\end{figure}

\section{Discussion and conclusions \label{sec:con}}

In this work, we developed a single-zone model to explain the
emission of PKS~1510-089 during the flare observed in March 2009. 
The distinct feature of models is 
 the low-energy component produced by the
synchrotron radiation and the high-energy component produced by the
same population of ultra-relativistic electrons via an IC process.  It
has been confirmed that the Comptonization of photons coming from BLR alone
cannot explain the VHE emission due to the KN effect, as anticipated
by \cite{2005MNRAS.363..954M};  VHE emission is accurately  explained by
the Comptonization of photons coming from the DT.

The recent attempt to model the observed spectrum of PKS~1510-089
was undertaken by \citet{2012ApJ...760...69N}, who analyzed the 2011
low state of the object.  They used the data obtained with the {\it
  Herschel} satellite to constrain the theoretical models and
concluded that a multi-zone emission model is necessary to explain the
spectral properties of PKS~1510-089.  

The fast optical flares observed in 2009 were
significantly brighter and more strongly polarized than those
observed in 2011.

\cite{2010ApJ...721.1425A} modeled the flares observed in 2009, 
but were unable to reproduce VHE emission due to 
assumption of BLR responsible for all HE emission.

The optical flaring activity in 2009 and 2011 might have been accompanied by
an increase in the magnetic field.  This is supported by the
observation of a significant increase in the degree of optical
polarization \citep{2011PASJ...63..489S} and observation of the
emergence of superluminal knot with VLBA at $7\,$mm
\citep{2010arXiv1002.0806M} during the flare in 2009.  This behavior
of the low-energy component was not observed in 2011 when the  {\it
  Herschel} data were taken.

The source PKS~1510-089 was detected with the H.E.S.S.\ system in March -- April
2009 during the high state in the HE and optical domains.  The
observations revealed a VHE emission up to $400\,$GeV.
\cite{2010ApJ...721.1425A} model the emission for energies below
$100\,$GeV, without predictions for the VHE emission.

Our model provides interesting predictions for variability pattern
observed at different wavelengths;
for example, the HE and X-ray emission 
does not have to be correlated
because the emission in these energy ranges is produced 
by different components.
This lack of correlation between X-ray and HE was reported in 
the multiwavelength data presented by \cite{2010ApJ...721.1425A},
who also report  a positive correlation between 
the HE and optical band.
The last correlation is a signature of a leptonic single-zone scenario
in which optical and $\gamma$-ray emission is produced by the same population of electrons.
Possible lack of correlation between HE and VHE 
can also be explained as due to different production mechanisms.

\cite{2010arXiv1002.0806M} demonstrated that the HE emission from the
jet of PKS~1510-089 is quite complex.  The emission arises from
different regions and probably multiple emission mechanisms are
involved.  Such a complicated behavior may be a result of different
locations of the blazar zone during different flares.  A complicated interplay
between the internal absorption of different spectral components in the HE
and VHE ranges may lead to large differences in the observed
correlation or lack of correlation at different wavelengths.

The absorption of the HE and the VHE photons in the blazar itself has also been  investigated.
Figure~\ref{fig:VHEsed} shows the blazar model with and without internal and external absorption corrections.
Strong absorption of the VHE photons by the
BLR photon field is avoided by locating the blazar zone outside the
BLR.  The absorption by photons from the DT and EBL absorption become
significant only for photons with energies above $400\,$GeV. Such a
highly energetic emission was not observed in the case of
PKS~1510-089.

The EBL models given by \cite{2008A&A...487..837F}, \cite{2011MNRAS.410.2556D}, and \cite{2010A&A...515A..19K}
have been used to analyze the absorption of VHE $\gamma$ rays in  intergalactic space.
These models are not significantly distinguishable within the H.E.S.S. errorbars. 
For  comparison, Figure~\ref{fig:VHEsed} shows the blazar model 
 after internal absorption and external absorption using the models given by
\cite{2008A&A...487..837F}, \cite{2009MNRAS.399.1694G}, and \cite{2006Natur.440.1018A}.

 Very high energy emission can be a  common feature of FSRQs.  
Because of their high luminosity FSRQs can provide valuable targets for 
future VHE experiments.
The H.E.S.S.\ II, with its energy range from tens of GeV, will provide a
great opportunity to search for emission from other objects of this
class.

In  \cite{2013arXiv1304.8071H} the Fermi-HESS $\gamma$-ray spectrum was used to put constraints on the EBL level. In the model described here the assumption of a single-emission component for the entire $\gamma$-ray regime does not hold. However, if the VHE emission is measured while the source is in a low state in the {\it Fermi}-band, then it is plausible that the EC component  from the torus dominates the VHE as well as the HE bands. In such a scenario the method used in  \cite{2013arXiv1304.8071H} is still applicable.

\begin{acknowledgements}
We would like to thank the referee for the valuable comments.
We would like to thank Jeff Grube for help with the {\it Swift}/XRT data analysis.
This work was supported by the Polish Ministry of Science and Higher Education
under Grants No. DEC-2011/01/N/ST9/06007. 
Part of this work was supported by the French national program 
for high-energy astrophysics PNHE, and the French ANR project CosmoTeV.
\end{acknowledgements}

\bibliographystyle{aa}
\bibliography{pks1510_aa}

\begin{thebibliography}{52}
\expandafter\ifx\csname natexlab\endcsname\relax\def\natexlab#1{#1}\fi

\bibitem[{{Abdo} {et~al.}(2010){Abdo}, {Ackermann}, {Agudo}, {Ajello},
  {Allafort}, {Aller}, {Aller}, {Antolini}, {Arkharov}, {Axelsson}, \&
  et~al.}]{2010ApJ...721.1425A}
{Abdo}, A.~A., {Ackermann}, M., {Agudo}, I., {et~al.} 2010, \apj, 721, 1425

\bibitem[{{Abdo} {et~al.}(2011){Abdo}, {Ackermann}, {Ajello}, {Allafort},
  {Baldini}, {Ballet}, {Barbiellini}, {Bastieri}, {Bellazzini}, {Berenji},
  {Blandford}, {Bloom}, {Bonamente}, {Borgland}, {Bouvier}, {Bregeon},
  {Brigida}, {Bruel}, {Buehler}, {Buson}, {Caliandro}, {Cameron}, {Caraveo},
  {Casandjian}, {Cavazzuti}, {Cecchi}, {Charles}, {Chekhtman}, {Cheung},
  {Chiang}, {Ciprini}, {Claus}, {Conrad}, {Cutini}, {D'Ammando}, {de Angelis},
  {de Palma}, {Dermer}, {Digel}, {Silva}, {Drell}, {Dubois}, {Dumora},
  {Escande}, {Favuzzi}, {Fegan}, {Ferrara}, {Fortin}, {Fukazawa}, {Fusco},
  {Gargano}, {Gasparrini}, {Gehrels}, {Germani}, {Giglietto}, {Giommi},
  {Giordano}, {Giroletti}, {Glanzman}, {Godfrey}, {Grenier}, {Grove},
  {Guiriec}, {Hadasch}, {Hayashida}, {Hays}, {Horan}, {Itoh},
  {J{\'o}hannesson}, {Johnson}, {Kamae}, {Katagiri}, {Kataoka},
  {Kn{\"o}dlseder}, {Kuss}, {Lande}, {Larsson}, {Latronico}, {Lee}, {Longo},
  {Loparco}, {Lott}, {Lovellette}, {Lubrano}, {Madejski}, {Makeev},
  {Mazziotta}, {McConville}, {McEnery}, {Michelson}, {Mitthumsiri}, {Mizuno},
  {Moiseev}, {Monte}, {Monzani}, {Morselli}, {Moskalenko}, {Murgia},
  {Naumann-Godo}, {Nishino}, {Nolan}, {Norris}, {Nuss}, {Ohsugi}, {Okumura},
  {Orlando}, {Ormes}, {Paneque}, {Pelassa}, {Pesce-Rollins}, {Pierbattista},
  {Piron}, {Porter}, {Rain{\`o}}, {Rando}, {Razzaque}, {Reimer}, {Reimer},
  {Ritz}, {Roth}, {Sadrozinski}, {Sanchez}, {Scargle}, {Schalk}, {Sgr{\`o}},
  {Siskind}, {Smith}, {Spandre}, {Spinelli}, {Strickman}, {Takahashi},
  {Takahashi}, {Tanaka}, {Tanaka}, {Thayer}, {Thayer}, {Thompson}, {Tibaldo},
  {Torres}, {Tosti}, {Tramacere}, {Troja}, {Vandenbroucke}, {Vasileiou},
  {Vianello}, {Vilchez}, {Vitale}, {Waite}, {Wang}, {Winer}, {Wood}, {Yang}, \&
  {Ziegler}}]{2011ApJ...733L..26A}
{Abdo}, A.~A., {Ackermann}, M., {Ajello}, M., {et~al.} 2011, \apjl, 733, L26

\bibitem[{{Abdo} {et~al.}(2009){Abdo}, {Ackermann}, {Ajello}, {Atwood},
  {Axelsson}, {Baldini}, {Ballet}, {Barbiellini}, {Bastieri}, {Battelino},
  {Baughman}, {Bechtol}, {Bellazzini}, {Berenji}, {Blandford}, {Bloom},
  {Bonamente}, {Borgland}, {Bouvier}, {Bregeon}, {Brez}, {Brigida}, {Bruel},
  {Burnett}, {Caliandro}, {Cameron}, {Caraveo}, {Casandjian}, {Cavazzuti},
  {Cecchi}, {Charles}, {Chaty}, {Chekhtman}, {Cheung}, {Chiang}, {Ciprini},
  {Claus}, {Cohen-Tanugi}, {Cominsky}, {Conrad}, {Costamante}, {Cutini},
  {Dermer}, {de Angelis}, {de Palma}, {Digel}, {Silva}, {Donato}, {Drell},
  {Dubois}, {Dumora}, {Farnier}, {Favuzzi}, {Focke}, {Foschini}, {Frailis},
  {Fuhrmann}, {Fukazawa}, {Funk}, {Fusco}, {Gargano}, {Gasparrini}, {Gehrels},
  {Germani}, {Giebels}, {Giglietto}, {Giommi}, {Giordano}, {Glanzman},
  {Godfrey}, {Grenier}, {Grondin}, {Grove}, {Guillemot}, {Guiriec}, {Hanabata},
  {Harding}, {Hartman}, {Hayashida}, {Hays}, {Hughes}, {J{\'o}hannesson},
  {Johnson}, {Johnson}, {Johnson}, {Kamae}, {Katagiri}, {Kataoka}, {Kawai},
  {Kerr}, {Kn{\"o}dlseder}, {Kocian}, {Kuehn}, {Kuss}, {Latronico}, {Lee},
  {Lemoine-Goumard}, {Longo}, {Loparco}, {Lott}, {Lovellette}, {Lubrano},
  {Madejski}, {Makeev}, {Massaro}, {Mazziotta}, {McEnery}, {McGlynn}, {Meurer},
  {Michelson}, {Mitthumsiri}, {Mizuno}, {Moiseev}, {Monte}, {Monzani},
  {Morselli}, {Moskalenko}, {Murgia}, {Nolan}, {Norris}, {Nuss}, {Ohsugi},
  {Omodei}, {Orlando}, {Ormes}, {Paneque}, {Panetta}, {Parent}, {Pelassa},
  {Pepe}, {Pesce-Rollins}, {Piron}, {Porter}, {Rain{\`o}}, {Rando}, {Razzano},
  {Reimer}, {Reimer}, {Reposeur}, {Reyes}, {Ritz}, {Rochester}, {Rodriguez},
  {Rahoui}, {Ryde}, {Sadrozinski}, {Sambruna}, {Sanchez}, {Sander},
  {Parkinson}, {Sgr{\`o}}, {Shaw}, {Smith}, {Smith}, {Spandre}, {Spinelli},
  {Starck}, {Strickman}, {Suson}, {Tajima}, {Takahashi}, {Takahashi}, {Tanaka},
  {Thayer}, {Thayer}, {Thompson}, {Tibaldo}, {Torres}, {Tosti}, {Tramacere},
  {Uchiyama}, {Usher}, {Vilchez}, {Villata}, {Vitale}, {Waite}, {Winer},
  {Wood}, {Ylinen}, {Zensus}, \& {Ziegler}}]{2009ApJ...699..817A}
{Abdo}, A.~A., {Ackermann}, M., {Ajello}, M., {et~al.} 2009, \apj, 699, 817

\bibitem[{{Abramowski} {et~al.}(2012){Abramowski}, {Acero}, {Aharonian},
  {Akhperjanian}, {Anton}, {Balenderan}, {Balzer}, {Barnacka}, {Becherini},
  {Becker Tjus}, {Bernl{\"o}hr}, {Birsin}, {Biteau}, {Bochow}, {Boisson},
  {Bolmont}, {Bordas}, {Brucker}, {Brun}, {Brun}, {Bulik}, {Carrigan},
  {Casanova}, {Cerruti}, {Chadwick}, {Charbonnier}, {Chaves}, {Cheesebrough},
  {Cologna}, {Conrad}, {Couturier}, {Dalton}, {Daniel}, {Davids}, {Degrange},
  {Deil}, {deWilt}, {Dickinson}, {Djannati-Ata{\"i}}, {Domainko}, {O'C.~Drury},
  {Dubus}, {Dutson}, {Dyks}, {Dyrda}, {Egberts}, {Eger}, {Espigat}, {Fallon},
  {Farnier}, {Fegan}, {Feinstein}, {Fernandes}, {Fernandez}, {Fiasson},
  {Fontaine}, {F{\"o}rster}, {F{\"u}{\ss}ling}, {Gajdus}, {Gallant},
  {Garrigoux}, {Gast}, {Giebels}, {Glicenstein}, {Gl{\"u}ck}, {G{\"o}ring},
  {Grondin}, {H{\"a}ffner}, {Hague}, {Hahn}, {Hampf}, {Harris}, {Heinz},
  {Heinzelmann}, {Henri}, {Hermann}, {Hillert}, {Hinton}, {Hofmann},
  {Hofverberg}, {Holler}, {Horns}, {Jacholkowska}, {Jahn}, {Jamrozy}, {Jung},
  {Kastendieck}, {Katarzy{\'n}ski}, {Katz}, {Kaufmann}, {Kh{\'e}lifi},
  {Klochkov}, {Klu{\'z}niak}, {Kneiske}, {Komin}, {Kosack}, {Kossakowski},
  {Krayzel}, {Laffon}, {Lamanna}, {Lenain}, {Lennarz}, {Lohse}, {Lopatin},
  {Lu}, {Marandon}, {Marcowith}, {Masbou}, {Maurin}, {Maxted}, {Mayer},
  {McComb}, {Medina}, {M{\'e}hault}, {Menzler}, {Moderski}, {Mohamed},
  {Moulin}, {Naumann}, {Naumann-Godo}, {de Naurois}, {Nedbal}, {Nguyen},
  {Niemiec}, {Nolan}, {Ohm}, {de O{\~n}a Wilhelmi}, {Opitz}, {Ostrowski},
  {Oya}, {Panter}, {Parsons}, {Paz Arribas}, {Pekeur}, {Pelletier}, {Perez},
  {Petrucci}, {Peyaud}, {Pita}, {P{\"u}hlhofer}, {Punch}, {Quirrenbach},
  {Raue}, {Reimer}, {Reimer}, {Renaud}, {de los Reyes}, {Rieger}, {Ripken},
  {Rob}, {Rosier-Lees}, {Rowell}, {Rudak}, {Rulten}, {Sahakian}, {Sanchez},
  {Santangelo}, {Schlickeiser}, {Schulz}, {Schwanke}, {Schwarzburg},
  {Schwemmer}, {Sheidaei}, {Skilton}, {Sol}, {Spengler}, {Stawarz},
  {Steenkamp}, {Stegmann}, {Stinzing}, {Stycz}, {Sushch}, {Szostek},
  {Tavernet}, {Terrier}, {Tluczykont}, {Valerius}, {van Eldik}, {Vasileiadis},
  {Venter}, {Viana}, {Vincent}, {V{\"o}lk}, {Volpe}, {Vorobiov}, {Vorster},
  {Wagner}, {Ward}, {White}, {Wierzcholska}, {Wouters}, {Zacharias}, {Zajczyk},
  {Zdziarski}, {Zech}, \& {Zechlin}}]{2012arXiv1212.3409H}
{Abramowski}, A., {Acero}, F., {Aharonian}, F., {et~al.} 2012, ArXiv e-prints

\bibitem[{{Aharonian} {et~al.}(2006){Aharonian}, {Akhperjanian}, {Bazer-Bachi},
  {Beilicke}, {Benbow}, {Berge}, {Bernl{\"o}hr}, {Boisson}, {Bolz}, {Borrel},
  {Braun}, {Breitling}, {Brown}, {Chadwick}, {Chounet}, {Cornils},
  {Costamante}, {Degrange}, {Dickinson}, {Djannati-Ata{\"i}}, {Drury}, {Dubus},
  {Emmanoulopoulos}, {Espigat}, {Feinstein}, {Fontaine}, {Fuchs}, {Funk},
  {Gallant}, {Giebels}, {Gillessen}, {Glicenstein}, {Goret}, {Hadjichristidis},
  {Hauser}, {Hauser}, {Heinzelmann}, {Henri}, {Hermann}, {Hinton}, {Hofmann},
  {Holleran}, {Horns}, {Jacholkowska}, {de Jager}, {Kh{\'e}lifi}, {Klages},
  {Komin}, {Konopelko}, {Latham}, {Le Gallou}, {Lemi{\`e}re},
  {Lemoine-Goumard}, {Leroy}, {Lohse}, {Martin}, {Martineau-Huynh},
  {Marcowith}, {Masterson}, {McComb}, {de Naurois}, {Nolan}, {Noutsos},
  {Orford}, {Osborne}, {Ouchrif}, {Panter}, {Pelletier}, {Pita},
  {P{\"u}hlhofer}, {Punch}, {Raubenheimer}, {Raue}, {Raux}, {Rayner}, {Reimer},
  {Reimer}, {Ripken}, {Rob}, {Rolland}, {Rowell}, {Sahakian}, {Saug{\'e}},
  {Schlenker}, {Schlickeiser}, {Schuster}, {Schwanke}, {Siewert}, {Sol},
  {Spangler}, {Steenkamp}, {Stegmann}, {Tavernet}, {Terrier}, {Th{\'e}oret},
  {Tluczykont}, {van Eldik}, {Vasileiadis}, {Venter}, {Vincent}, {V{\"o}lk}, \&
  {Wagner}}]{2006Natur.440.1018A}
{Aharonian}, F., {Akhperjanian}, A.~G., {Bazer-Bachi}, A.~R., {et~al.} 2006,
  \nat, 440, 1018

\bibitem[{{Aleksi{\'c}} {et~al.}(2011{\natexlab{a}}){Aleksi{\'c}}, {Antonelli},
  {Antoranz}, {Backes}, {Barrio}, {Bastieri}, {Becerra Gonz{\'a}lez},
  {Bednarek}, {Berdyugin}, {Berger}, {Bernardini}, {Biland}, {Blanch}, {Bock},
  {Boller}, {Bonnoli}, {Borla Tridon}, {Braun}, {Bretz}, {Ca{\~n}ellas},
  {Carmona}, {Carosi}, {Colin}, {Colombo}, {Contreras}, {Cortina}, {Cossio},
  {Covino}, {Dazzi}, {de Angelis}, {de Cea Del Pozo}, {de Lotto}, {Delgado
  Mendez}, {Diago Ortega}, {Doert}, {Dom{\'{\i}}nguez}, {Dominis Prester},
  {Dorner}, {Doro}, {Elsaesser}, {Ferenc}, {Fonseca}, {Font}, {Fruck},
  {Garc{\'{\i}}a L{\'o}pez}, {Garczarczyk}, {Garrido}, {Giavitto},
  {Godinovi{\'c}}, {Hadasch}, {H{\"a}fner}, {Herrero}, {Hildebrand}, {Hose},
  {Hrupec}, {Huber}, {Jogler}, {Klepser}, {Kr{\"a}henb{\"u}hl}, {Krause}, {La
  Barbera}, {Lelas}, {Leonardo}, {Lindfors}, {Lombardi}, {L{\'o}pez}, {Lorenz},
  {Majumdar}, {Makariev}, {Maneva}, {Mankuzhiyil}, {Mannheim}, {Maraschi},
  {Mariotti}, {Mart{\'{\i}}nez}, {Mazin}, {Meucci}, {Miranda}, {Mirzoyan},
  {Miyamoto}, {Mold{\'o}n}, {Moralejo}, {Nieto}, {Nilsson}, {Orito}, {Oya},
  {Paoletti}, {Pardo}, {Paredes}, {Partini}, {Pasanen}, {Pauss},
  {Perez-Torres}, {Persic}, {Peruzzo}, {Pilia}, {Pochon}, {Prada}, {Prada
  Moroni}, {Prandini}, {Puljak}, {Reichardt}, {Reinthal}, {Rhode}, {Rib{\'o}},
  {Rico}, {R{\"u}gamer}, {R{\"u}ger}, {Saggion}, {Saito}, {Saito}, {Salvati},
  {Satalecka}, {Scalzotto}, {Scapin}, {Schultz}, {Schweizer}, {Shayduk},
  {Shore}, {Sillanp{\"a}{\"a}}, {Sitarek}, {Sobczynska}, {Spanier}, {Spiro},
  {Stamerra}, {Steinke}, {Storz}, {Strah}, {Suri{\'c}}, {Takalo}, {Tavecchio},
  {Temnikov}, {Terzi{\'c}}, {Tescaro}, {Teshima}, {Thom}, {Tibolla}, {Torres},
  {Treves}, {Vankov}, {Vogler}, {Wagner}, {Weitzel}, {Zabalza}, {Zandanel}, \&
  {Zanin}}]{2011A&A...530A...4A}
{Aleksi{\'c}}, J., {Antonelli}, L.~A., {Antoranz}, P., {et~al.}
  2011{\natexlab{a}}, \aap, 530, A4

\bibitem[{{Aleksi{\'c}} {et~al.}(2011{\natexlab{b}}){Aleksi{\'c}}, {Antonelli},
  {Antoranz}, {Backes}, {Barrio}, {Bastieri}, {Becerra Gonz{\'a}lez},
  {Bednarek}, {Berdyugin}, {Berger}, {Bernardini}, {Biland}, {Blanch}, {Bock},
  {Boller}, {Bonnoli}, {Borla Tridon}, {Braun}, {Bretz}, {Ca{\~n}ellas},
  {Carmona}, {Carosi}, {Colin}, {Colombo}, {Contreras}, {Cortina}, {Cossio},
  {Covino}, {Dazzi}, {De Angelis}, {De Cea del Pozo}, {De Lotto}, {Delgado
  Mendez}, {Diago Ortega}, {Doert}, {Dom{\'{\i}}nguez}, {Dominis Prester},
  {Dorner}, {Doro}, {Elsaesser}, {Ferenc}, {Fonseca}, {Font}, {Fruck},
  {Garc{\'{\i}}a L{\'o}pez}, {Garczarczyk}, {Garrido}, {Giavitto},
  {Godinovi{\'c}}, {Hadasch}, {H{\"a}fner}, {Herrero}, {Hildebrand},
  {H{\"o}hne-M{\"o}nch}, {Hose}, {Hrupec}, {Huber}, {Jogler}, {Klepser},
  {Kr{\"a}henb{\"u}hl}, {Krause}, {La Barbera}, {Lelas}, {Leonardo},
  {Lindfors}, {Lombardi}, {L{\'o}pez}, {Lorenz}, {Makariev}, {Maneva},
  {Mankuzhiyil}, {Mannheim}, {Maraschi}, {Mariotti}, {Mart{\'{\i}}nez},
  {Mazin}, {Meucci}, {Miranda}, {Mirzoyan}, {Miyamoto}, {Mold{\'o}n},
  {Moralejo}, {Nieto}, {Nilsson}, {Orito}, {Oya}, {Paneque}, {Paoletti},
  {Pardo}, {Paredes}, {Partini}, {Pasanen}, {Pauss}, {Perez-Torres}, {Persic},
  {Peruzzo}, {Pilia}, {Pochon}, {Prada}, {Prada Moroni}, {Prandini}, {Puljak},
  {Reichardt}, {Reinthal}, {Rhode}, {Rib{\'o}}, {Rico}, {R{\"u}gamer},
  {Saggion}, {Saito}, {Saito}, {Salvati}, {Satalecka}, {Scalzotto}, {Scapin},
  {Schultz}, {Schweizer}, {Shayduk}, {Shore}, {Sillanp{\"a}{\"a}}, {Sitarek},
  {Sobczynska}, {Spanier}, {Spiro}, {Stamerra}, {Steinke}, {Storz}, {Strah},
  {Suri{\'c}}, {Takalo}, {Tavecchio}, {Temnikov}, {Terzi{\'c}}, {Tescaro},
  {Teshima}, {Thom}, {Tibolla}, {Torres}, {Treves}, {Vankov}, {Vogler},
  {Wagner}, {Weitzel}, {Zabalza}, {Zandanel}, {Zanin}, {MAGIC Collaboration},
  {Tanaka}, {Wood}, \& {Buson}}]{2011ApJ...730L...8A}
{Aleksi{\'c}}, J., {Antonelli}, L.~A., {Antoranz}, P., {et~al.}
  2011{\natexlab{b}}, \apjl, 730, L8

\bibitem[{{Atwood} {et~al.}(2009){Atwood}, {Abdo}, {Ackermann}, {Althouse},
  {Anderson}, {Axelsson}, {Baldini}, {Ballet}, {Band}, {Barbiellini}, \&
  et~al.}]{2009Atwood}
{Atwood}, W.~B., {Abdo}, A.~A., {Ackermann}, M., {et~al.} 2009, \apj, 697, 1071

\bibitem[{{Barnacka} \& {Moderski}(2009)}]{ABIntegral}
{Barnacka}, A. \& {Moderski}, R. 2009, in Very High Energy Phenomena in the
  Universe, ed. J.~{Dumarchez} \& J.~{Tran Thanh Van}, Vol.~1, 111

\bibitem[{{Bi} \& {Yuan}(2008)}]{2008arXiv0809.5124B}
{Bi}, X.-J. \& {Yuan}, Q. 2008, ArXiv e-prints

\bibitem[{{Blandford} \& {Rees}(1978)}]{1978bllo.conf..328B}
{Blandford}, R.~D. \& {Rees}, M.~J. 1978, in BL Lac Objects, ed. A.~M. {Wolfe},
  328--341

\bibitem[{{B{\l}a{\.z}ejowski} {et~al.}(2000){B{\l}a{\.z}ejowski}, {Sikora},
  {Moderski}, \& {Madejski}}]{2000ApJ...545..107B}
{B{\l}a{\.z}ejowski}, M., {Sikora}, M., {Moderski}, R., \& {Madejski}, G.~M.
  2000, \apj, 545, 107

\bibitem[{{D'Ammando} {et~al.}(2009){D'Ammando}, {Vercellone}, {Tavani},
  {Pucella}, {Chen}, {Giuliani}, {Bulgarelli}, {Piano}, {Vittorini}, {Costa},
  {Feroci}, {Donnarumma}, {Pacciani}, {Del Monte}, {Lazzarotto}, {Soffitta},
  {Evangelista}, {Lapshov}, {Rapisarda}, {Argan}, {Trois}, {de Paris},
  {Marisaldi}, {Gianotti}, {Trifoglio}, {Di Cocco}, {Labanti}, {Fuschino},
  {Galli}, {Caraveo}, {Mereghetti}, {Perotti}, {Fiorini}, {Zambra},
  {Pellizzoni}, {Pilia}, {Barbiellini}, {Longo}, {Moretti}, {Vallazza},
  {Picozza}, {Morselli}, {Sabatini}, {Prest}, {Lipari}, {Zanello}, {Cattaneo},
  {Pittori}, {Verrecchia}, {Santolamazza}, {Colafrancesco}, {Giommi}, \&
  {Salotti}}]{2009ATel.1957....1D}
{D'Ammando}, F., {Vercellone}, S., {Tavani}, M., {et~al.} 2009, The
  Astronomer's Telegram, 1957, 1

\bibitem[{{Dom{\'{\i}}nguez} {et~al.}(2011){Dom{\'{\i}}nguez}, {Primack},
  {Rosario}, {Prada}, {Gilmore}, {Faber}, {Koo}, {Somerville},
  {P{\'e}rez-Torres}, {P{\'e}rez-Gonz{\'a}lez}, {Huang}, {Davis},
  {Guhathakurta}, {Barmby}, {Conselice}, {Lozano}, {Newman}, \&
  {Cooper}}]{2011MNRAS.410.2556D}
{Dom{\'{\i}}nguez}, A., {Primack}, J.~R., {Rosario}, D.~J., {et~al.} 2011,
  \mnras, 410, 2556

\bibitem[{{Donea} \& {Protheroe}(2003)}]{2003APh....18..377D}
{Donea}, A.-C. \& {Protheroe}, R.~J. 2003, Astroparticle Physics, 18, 377

\bibitem[{{Errando} {et~al.}(2012){Errando}, {}, \& {for the VERITAS
  Collaboration}}]{2012arXiv1205.0068E}
{Errando}, M., {}, \& {for the VERITAS Collaboration}. 2012, ArXiv e-prints

\bibitem[{{Evans} {et~al.}(2009){Evans}, {Beardmore}, {Page}, {Osborne},
  {O'Brien}, {Willingale}, {Starling}, {Burrows}, {Godet}, {Vetere}, {Racusin},
  {Goad}, {Wiersema}, {Angelini}, {Capalbi}, {Chincarini}, {Gehrels}, {Kennea},
  {Margutti}, {Morris}, {Mountford}, {Pagani}, {Perri}, {Romano}, \&
  {Tanvir}}]{2009MNRAS.397.1177E}
{Evans}, P.~A., {Beardmore}, A.~P., {Page}, K.~L., {et~al.} 2009, \mnras, 397,
  1177

\bibitem[{{Franceschini} {et~al.}(2008){Franceschini}, {Rodighiero}, \&
  {Vaccari}}]{2008A&A...487..837F}
{Franceschini}, A., {Rodighiero}, G., \& {Vaccari}, M. 2008, \aap, 487, 837

\bibitem[{{Ghisellini} \& {Tavecchio}(2009)}]{2009MNRAS.397..985G}
{Ghisellini}, G. \& {Tavecchio}, F. 2009, \mnras, 397, 985

\bibitem[{{Gilmore} {et~al.}(2009){Gilmore}, {Madau}, {Primack}, {Somerville},
  \& {Haardt}}]{2009MNRAS.399.1694G}
{Gilmore}, R.~C., {Madau}, P., {Primack}, J.~R., {Somerville}, R.~S., \&
  {Haardt}, F. 2009, \mnras, 399, 1694

\bibitem[{{Gould} \& {Schr{\'e}der}(1967)}]{1967PhRv..155.1404G}
{Gould}, R.~J. \& {Schr{\'e}der}, G.~P. 1967, Physical Review, 155, 1404

\bibitem[{{Hartman} {et~al.}(1999){Hartman}, {Bertsch}, {Bloom}, {Chen},
  {Deines-Jones}, {Esposito}, {Fichtel}, {Friedlander}, {Hunter}, {McDonald},
  {Sreekumar}, {Thompson}, {Jones}, {Lin}, {Michelson}, {Nolan}, {Tompkins},
  {Kanbach}, {Mayer-Hasselwander}, {M{\"u}cke}, {Pohl}, {Reimer}, {Kniffen},
  {Schneid}, {von Montigny}, {Mukherjee}, \& {Dingus}}]{1999ApJS..123...79H}
{Hartman}, R.~C., {Bertsch}, D.~L., {Bloom}, S.~D., {et~al.} 1999, \apjs, 123,
  79

\bibitem[{{Hauser} {et~al.}(2011){Hauser}, {Lenain}, {Wagner}, \&
  {Hagen}}]{2011ATel.3509....1H}
{Hauser}, M., {Lenain}, J.~P., {Wagner}, S., \& {Hagen}, H. 2011, The
  Astronomer's Telegram, 3509, 1

\bibitem[{{Hauser} \& {Dwek}(2001)}]{2001ARA&A..39..249H}
{Hauser}, M.~G. \& {Dwek}, E. 2001, \araa, 39, 249

\bibitem[{{Hayashida} {et~al.}(2012){Hayashida}, {Madejski}, {Nalewajko},
  {Sikora}, {Wehrle}, {Ogle}, {Collmar}, {Larsson}, {Fukazawa}, {Itoh},
  {Chiang}, {Stawarz}, {Blandford}, {Richards}, {Max-Moerbeck}, {Readhead},
  {Buehler}, {Cavazzuti}, {Ciprini}, {Gehrels}, {Reimer}, {Szostek}, {Tanaka},
  {Tosti}, {Uchiyama}, {Kawabata}, {Kino}, {Sakimoto}, {Sasada}, {Sato},
  {Uemura}, {Yamanaka}, {Greiner}, {Kruehler}, {Rossi}, {Macquart}, {Bock},
  {Villata}, {Raiteri}, {Agudo}, {Aller}, {Aller}, {Arkharov}, {Bach},
  {Ben{\'{\i}}tez}, {Berdyugin}, {Blinov}, {Blumenthal}, {B{\"o}ttcher},
  {Buemi}, {Carosati}, {Chen}, {Di Paola}, {Dolci}, {Efimova}, {Forn{\'e}},
  {G{\'o}mez}, {Gurwell}, {Heidt}, {Hiriart}, {Jordan}, {Jorstad}, {Joshi},
  {Kimeridze}, {Konstantinova}, {Kopatskaya}, {Koptelova}, {Kurtanidze},
  {L{\"a}hteenm{\"a}ki}, {Lamerato}, {Larionov}, {Larionova}, {Larionova},
  {Leto}, {Lindfors}, {Marscher}, {McHardy}, {Molina}, {Morozova},
  {Nikolashvili}, {Nilsson}, {Reinthal}, {Roustazadeh}, {Sakamoto}, {Sigua},
  {Sillanp{\"a}{\"a}}, {Takalo}, {Tammi}, {Taylor}, {Tornikoski}, {Trigilio},
  {Troitsky}, \& {Umana}}]{2012ApJ...754..114H}
{Hayashida}, M., {Madejski}, G.~M., {Nalewajko}, K., {et~al.} 2012, \apj, 754,
  114

\bibitem[{{H.E.S.S.~Collaboration} {et~al.}(2013){H.E.S.S.~Collaboration},
  {Abramowski}, {Acero}, {Aharonian}, {Akhperjanian}, {Anton}, {Balenderan},
  {Balzer}, {Barnacka}, {Becherini}, {Becker Tjus}, {Behera}, {Bernl{\"o}hr},
  {Birsin}, {Biteau}, {Bochow}, {Boisson}, {Bolmont}, {Bordas}, {Brucker},
  {Brun}, {Brun}, {Bulik}, {Carrigan}, {Casanova}, {Cerruti}, {Chadwick},
  {Chaves}, {Cheesebrough}, {Colafrancesco}, {Cologna}, {Conrad}, {Couturier},
  {Dalton}, {Daniel}, {Davids}, {Degrange}, {Deil}, {deWilt}, {Dickinson},
  {Djannati-Ata{\"i}}, {Domainko}, {O'C.~Drury}, {Dubus}, {Dutson}, {Dyks},
  {Dyrda}, {Egberts}, {Eger}, {Espigat}, {Fallon}, {Farnier}, {Fegan},
  {Feinstein}, {Fernandes}, {Fernandez}, {Fiasson}, {Fontaine}, {F{\"o}rster},
  {F{\"u}{\ss}ling}, {Gajdus}, {Gallant}, {Garrigoux}, {Gast}, {Giebels},
  {Glicenstein}, {Gl{\"u}ck}, {G{\"o}ring}, {Grondin}, {Grudzi{\'n}ska},
  {H{\"a}ffner}, {Hague}, {Hahn}, {Hampf}, {Harris}, {Hauser}, {Heinz},
  {Heinzelmann}, {Henri}, {Hermann}, {Hillert}, {Hinton}, {Hofmann},
  {Hofverberg}, {Holler}, {Horns}, {Jacholkowska}, {Jahn}, {Jamrozy}, {Jung},
  {Kastendieck}, {Katarzy{\'n}ski}, {Katz}, {Kaufmann}, {Kh{\'e}lifi},
  {Klepser}, {Klochkov}, {Klu{\'z}niak}, {Kneiske}, {Kolitzus}, {Komin},
  {Kosack}, {Kossakowski}, {Krayzel}, {Kr{\"u}ger}, {Laffon}, {Lamanna},
  {Lefaucheur}, {Lemoine-Goumard}, {Lenain}, {Lennarz}, {Lohse}, {Lopatin},
  {Lu}, {Marandon}, {Marcowith}, {Masbou}, {Maurin}, {Maxted}, {Mayer},
  {McComb}, {Medina}, {M{\'e}hault}, {Menzler}, {Moderski}, {Mohamed},
  {Moulin}, {Naumann}, {Naumann-Godo}, {de Naurois}, {Nedbal}, {Nguyen},
  {Niemiec}, {Nolan}, {Ohm}, {de O{\~n}a Wilhelmi}, {Opitz}, {Ostrowski},
  {Oya}, {Panter}, {Parsons}, {Paz Arribas}, {Pekeur}, {Pelletier}, {Perez},
  {Petrucci}, {Peyaud}, {Pita}, {P{\"u}hlhofer}, {Punch}, {Quirrenbach},
  {Raab}, {Raue}, {Reimer}, {Reimer}, {Renaud}, {de los Reyes}, {Rieger},
  {Ripken}, {Rob}, {Rosier-Lees}, {Rowell}, {Rudak}, {Rulten}, {Sahakian},
  {Sanchez}, {Santangelo}, {Schlickeiser}, {Schulz}, {Schwanke}, {Schwarzburg},
  {Schwemmer}, {Sheidaei}, {Skilton}, {Sol}, {Spengler}, {Stawarz},
  {Steenkamp}, {Stegmann}, {Stinzing}, {Stycz}, {Sushch}, {Szostek},
  {Tavernet}, {Terrier}, {Tluczykont}, {Trichard}, {Valerius}, {van Eldik},
  {Vasileiadis}, {Venter}, {Viana}, {Vincent}, {V{\"o}lk}, {Volpe}, {Vorobiov},
  {Vorster}, {Wagner}, {Ward}, {White}, {Wierzcholska}, {Wouters}, {Zacharias},
  {Zajczyk}, {Zdziarski}, {Zech}, \& {Zechlin}}]{2013arXiv1304.8071H}
{H.E.S.S.~Collaboration}, {Abramowski}, A., {Acero}, F., {et~al.} 2013, \aap,
  554, A107

\bibitem[{{Homan} {et~al.}(2001){Homan}, {Ojha}, {Wardle}, {Roberts}, {Aller},
  {Aller}, \& {Hughes}}]{2001ApJ...549..840H}
{Homan}, D.~C., {Ojha}, R., {Wardle}, J.~F.~C., {et~al.} 2001, \apj, 549, 840

\bibitem[{{Homan} {et~al.}(2002){Homan}, {Wardle}, {Cheung}, {Roberts}, \&
  {Attridge}}]{2002ApJ...580..742H}
{Homan}, D.~C., {Wardle}, J.~F.~C., {Cheung}, C.~C., {Roberts}, D.~H., \&
  {Attridge}, J.~M. 2002, \apj, 580, 742

\bibitem[{{Jorstad} {et~al.}(2005){Jorstad}, {Marscher}, {Lister}, {Stirling},
  {Cawthorne}, {Gear}, {G{\'o}mez}, {Stevens}, {Smith}, {Forster}, \&
  {Robson}}]{2005AJ....130.1418J}
{Jorstad}, S.~G., {Marscher}, A.~P., {Lister}, M.~L., {et~al.} 2005, \aj, 130,
  1418

\bibitem[{{Kalberla} {et~al.}(2005){Kalberla}, {Burton}, {Hartmann}, {Arnal},
  {Bajaja}, {Morras}, \& {P{\"o}ppel}}]{2005A&A...440..775K}
{Kalberla}, P.~M.~W., {Burton}, W.~B., {Hartmann}, D., {et~al.} 2005, \aap,
  440, 775

\bibitem[{{Kataoka} {et~al.}(2008){Kataoka}, {Madejski}, {Sikora}, {Roming},
  {Chester}, {Grupe}, {Tsubuku}, {Sato}, {Kawai}, {Tosti}, {Impiombato},
  {Kovalev}, {Kovalev}, {Edwards}, {Wagner}, {Moderski}, {Stawarz},
  {Takahashi}, \& {Watanabe}}]{2008ApJ...672..787K}
{Kataoka}, J., {Madejski}, G., {Sikora}, M., {et~al.} 2008, \apj, 672, 787

\bibitem[{{Kneiske} \& {Dole}(2010)}]{2010A&A...515A..19K}
{Kneiske}, T.~M. \& {Dole}, H. 2010, \aap, 515, A19

\bibitem[{{Liu} \& {Bai}(2006)}]{2006ApJ...653.1089L}
{Liu}, H.~T. \& {Bai}, J.~M. 2006, \apj, 653, 1089

\bibitem[{{Marscher} {et~al.}(2010){Marscher}, {Jorstad}, {D'Arcangelo},
  {Bhattarai}, {Taylor}, {Olmstead}, {Manne-Nicholas}, {Larionov},
  {Hagen-Thorn}, {Konstantinova}, {Larionova}, {Larionova}, {Melnichuk},
  {Blinov}, {Kopatskaya}, {Troitsky}, {Agudo}, {G{\'o}mez}, {Roca-Sogorb},
  {Smith}, {Schmidt}, {Kurtanidze}, {Nikolashvili}, {Kimeridze}, \&
  {Sigua}}]{2010arXiv1002.0806M}
{Marscher}, A.~P., {Jorstad}, S.~G., {D'Arcangelo}, F.~D., {et~al.} 2010, ArXiv
  e-prints

\bibitem[{{Meyer} {et~al.}(2012){Meyer}, {Raue}, {Mazin}, \&
  {Horns}}]{2012A&A...542A..59M}
{Meyer}, M., {Raue}, M., {Mazin}, D., \& {Horns}, D. 2012, \aap, 542, A59

\bibitem[{{Moderski} {et~al.}(2003){Moderski}, {Sikora}, \&
  {B{\l}a{\.z}ejowski}}]{2003A&A...406..855M}
{Moderski}, R., {Sikora}, M., \& {B{\l}a{\.z}ejowski}, M. 2003, \aap, 406, 855

\bibitem[{{Moderski} {et~al.}(2005){Moderski}, {Sikora}, {Coppi}, \&
  {Aharonian}}]{2005MNRAS.363..954M}
{Moderski}, R., {Sikora}, M., {Coppi}, P.~S., \& {Aharonian}, F. 2005, \mnras,
  363, 954

\bibitem[{{Nalewajko} {et~al.}(2012){Nalewajko}, {Sikora}, {Madejski}, {Exter},
  {Szostek}, {Szczerba}, {Kidger}, \& {Lorente}}]{2012ApJ...760...69N}
{Nalewajko}, K., {Sikora}, M., {Madejski}, G.~M., {et~al.} 2012, \apj, 760, 69

\bibitem[{{Nolan} {et~al.}(2012){Nolan}, {Abdo}, {Ackermann}, {Ajello},
  {Allafort}, {Antolini}, {Atwood}, {Axelsson}, {Baldini}, {Ballet}, \&
  et~al.}]{2012ApJS..199...31N}
{Nolan}, P.~L., {Abdo}, A.~A., {Ackermann}, M., {et~al.} 2012, \apjs, 199, 31

\bibitem[{{Orienti} {et~al.}(2013){Orienti}, {Koyama}, {D'Ammando},
  {Giroletti}, {Kino}, {Nagai}, {Venturi}, {Dallacasa}, {Giovannini},
  {Angelakis}, {Fuhrmann}, {Hovatta}, {Max-Moerbeck}, {Schinzel}, {Akiyama},
  {Hada}, {Honma}, {Niinuma}, {Gasparrini}, {Krichbaum}, {Nestoras},
  {Readhead}, {Richards}, {Riquelme}, {Sievers}, {Ungerechts}, \&
  {Zensus}}]{2012arXiv1210.4319O}
{Orienti}, M., {Koyama}, S., {D'Ammando}, F., {et~al.} 2013, \mnras, 428, 2418

\bibitem[{{Pian} {et~al.}(2005){Pian}, {Falomo}, \&
  {Treves}}]{2005MNRAS.361..919P}
{Pian}, E., {Falomo}, R., \& {Treves}, A. 2005, \mnras, 361, 919

\bibitem[{{Poutanen} \& {Stern}(2010)}]{2010ApJ...717L.118P}
{Poutanen}, J. \& {Stern}, B. 2010, \apjl, 717, L118

\bibitem[{{Pucella} {et~al.}(2009){Pucella}, {D'Ammando}, {Tavani}, {Del
  Monte}, {Marisaldi}, {Gianotti}, {Trifoglio}, {Bulgarelli}, {Pittori},
  {Verrecchia}, {Vercellone}, {Chen}, {Giuliani}, {Piano}, {Vittorini},
  {Costa}, {Feroci}, {Donnarumma}, {Pacciani}, {Lazzarotto}, {Soffitta},
  {Evangelista}, {Lapshov}, {Rapisarda}, {Argan}, {Trois}, {de Paris}, {Di
  Cocco}, {Labanti}, {Fuschino}, {Galli}, {Caraveo}, {Mereghetti}, {Perotti},
  {Fiorini}, {Zambra}, {Pellizzoni}, {Pilia}, {Barbiellini}, {Longo},
  {Moretti}, {Vallazza}, {Picozza}, {Morselli}, {Sabatini}, {Prest}, {Lipari},
  {Zanello}, {Cattaneo}, {Santolamazza}, {Colafrancesco}, {Giommi}, \&
  {Salotti}}]{2009ATel.1968....1P}
{Pucella}, G., {D'Ammando}, F., {Tavani}, M., {et~al.} 2009, The Astronomer's
  Telegram, 1968, 1

\bibitem[{{Saito} {et~al.}(2013){Saito}, {Stawarz}, {Tanaka}, {Takahashi},
  {Madejski}, \& {D'Ammando}}]{2013ApJ...766L..11S}
{Saito}, S., {Stawarz}, {\L}., {Tanaka}, Y.~T., {et~al.} 2013, \apjl, 766, L11

\bibitem[{{Sasada} {et~al.}(2011){Sasada}, {Uemura}, {Fukazawa}, {Kawabata},
  {Ikejiri}, {Itoh}, {Yamanaka}, {Sakimoto}, {Ohsugi}, {Yoshida}, {Sato}, \&
  {Kino}}]{2011PASJ...63..489S}
{Sasada}, M., {Uemura}, M., {Fukazawa}, Y., {et~al.} 2011, \pasj, 63, 489

\bibitem[{{Sikora} {et~al.}(1994){Sikora}, {Begelman}, \&
  {Rees}}]{1994ApJ...421..153S}
{Sikora}, M., {Begelman}, M.~C., \& {Rees}, M.~J. 1994, \apj, 421, 153

\bibitem[{{Sikora} {et~al.}(2013){Sikora}, {Janiak}, {Nalewajko}, {Madejski},
  \& {Moderski}}]{2013ApJ...779...68S}
{Sikora}, M., {Janiak}, M., {Nalewajko}, K., {Madejski}, G.~M., \& {Moderski},
  R. 2013, \apj, 779, 68

\bibitem[{{Sikora} {et~al.}(2009){Sikora}, {Stawarz}, {Moderski}, {Nalewajko},
  \& {Madejski}}]{2009ApJ...704...38S}
{Sikora}, M., {Stawarz}, {\L}., {Moderski}, R., {Nalewajko}, K., \& {Madejski},
  G.~M. 2009, \apj, 704, 38

\bibitem[{{Vercellone} {et~al.}(2009){Vercellone}, {D'Ammando}, {Pucella},
  {Tavani}, {Donnarumma}, {Pacciani}, {Bulgarelli}, {Marisaldi}, {Gianotti},
  {Trifoglio}, {Pittori}, {Verrecchia}, {Chen}, {Giuliani}, {Piano},
  {Vittorini}, {Costa}, {Feroci}, {Del Monte}, {Lazzarotto}, {Soffitta},
  {Evangelista}, {Lapshov}, {Rapisarda}, {Argan}, {Trois}, {de Paris}, {Di
  Cocco}, {Labanti}, {Fuschino}, {Galli}, {Caraveo}, {Mereghetti}, {Perotti},
  {Fiorini}, {Zambra}, {Pellizzoni}, {Pilia}, {Barbiellini}, {Longo},
  {Moretti}, {Vallazza}, {Picozza}, {Morselli}, {Sabatini}, {Prest}, {Lipari},
  {Zanello}, {Cattaneo}, {Rappoldi}, {Santolamazza}, {Colafrancesco}, {Giommi},
  \& {Salotti}}]{2009ATel.1976....1V}
{Vercellone}, S., {D'Ammando}, F., {Pucella}, G., {et~al.} 2009, The
  Astronomer's Telegram, 1976, 1

\bibitem[{{Villata} {et~al.}(2009){Villata}, {Raiteri}, {Larionov},
  {Gorshanov}, {Konstantinova}, {Kopatskaya}, {Larionova}, {Chen}, {Koptelova},
  {Nilsson}, \& {Pasanen}}]{2009ATel.1988....1V}
{Villata}, M., {Raiteri}, C.~M., {Larionov}, V.~M., {et~al.} 2009, The
  Astronomer's Telegram, 1988, 1

\bibitem[{{Wagner} \& {H.E.S.S.~Collaboration}(2010)}]{2010HEAD...11.2706W}
{Wagner}, S.~J. \& {H.E.S.S.~Collaboration}. 2010, in AAS/High Energy
  Astrophysics Division, Vol.~11, AAS/High Energy Astrophysics Division \#11,
  \#27.06

\bibitem[{{Wardle} {et~al.}(2005){Wardle}, {Homan}, {Cheung}, \&
  {Roberts}}]{2005ASPC..340...67W}
{Wardle}, J.~F.~C., {Homan}, D.~C., {Cheung}, C.~C., \& {Roberts}, D.~H. 2005,
  in Astronomical Society of the Pacific Conference Series, Vol. 340, Future
  Directions in High Resolution Astronomy, ed. J.~{Romney} \& M.~{Reid}, 67

\end{thebibliography}

\end{document}